\documentclass[12pt]{article}
\usepackage{newtxtext,newtxmath}
\usepackage{graphicx}
\usepackage[letterpaper,margin=1in]{geometry}
\renewenvironment{abstract}
	{\quotation}
	{\endquotation}
\date{}

\makeatletter
\renewcommand{\fnum@figure}{\textbf{Figure \thefigure}}
\renewcommand{\fnum@table}{\textbf{Table \thetable}}
\makeatother
\usepackage{scicite}
\usepackage{url}

\def\scititle{
    CyberDiver: an untethered robotic impactor for water-entry experiments
}
\title{\bfseries \boldmath \scititle}

\author{
    John~T.~Antolik$^{1}$,
    Eli~A.~Silver$^{1}$,
    Jesse~L.~Belden$^{2}$,
    Daniel~M.~Harris$^{1\ast}$\and
    \small$^{1}$Center for Fluid Mechanics and School of Engineering, Brown University, Providence, RI 02912, USA.\and
    \small$^{2}$Naval Undersea Warfare Center Division Newport, Newport, RI 02841, USA.\and
    \small$^\ast$Corresponding author. Email: daniel\_harris3@brown.edu\and
}

\begin{document} 

\maketitle

\begin{abstract} \bfseries \boldmath

We present the CyberDiver, an untethered robotic impactor capable of actively modulating the fluid physics during high-speed water entry. First, we utilize the CyberDiver to extend our understanding of the water entry of passively flexible systems, designing a high-bandwidth controller that enables the CyberDiver to operate as a cyber-physical system that permits an arbitrary programmable structural coupling to be experimentally tested. Onboard sensors record the body acceleration during impact and reveal that the introduction of damping or a nonlinear force-versus-displacement structural law can significantly reduce impact loading as compared to a linear elastic case, with implications for damage mitigation in aerospace and naval applications. Next, by operating the CyberDiver in a displacement control mode, we demonstrate that the splash size can be dramatically altered depending on the parameters of an active maneuver, laying a groundwork for better understanding the techniques of human competitive divers. 

\end{abstract}

\section*{Introduction}

An understanding of the hydrodynamics during impact is necessary to prevent damage to engineered structures that operate at the air-water interface such as ships, seaplanes, projectiles, or re-entry capsules \cite{li2025review, mcgehee1959nasa, seddon2006aerospace}. Water entry poses a risk of injury to biological divers including humans and sea birds as well \cite{pandey2022slamming}. Recently there is interest in the construction of robots which operate at the air-water interface and perform water entry or exit at both a small scale, inspired by biological systems, \cite{chen2017microrobot, li2022airwaterdrone} and at larger scales for autonomous underwater exploration and monitoring \cite{wang2021auvnumerical, yan2018auvexperimental}. Blunt objects experience intense impact loading as they initially enter the water surface during the so-called slamming phase, an observation reconciled theoretically in early works by von Karman \cite{von1929impact} and Wagner \cite{wagner1932stoss}. The impact loading occurs due to the effect of fluid added mass: as the body impacts the surface it must suddenly accelerate an appreciable quantity of water, thereby undergoing a substantial transfer of momentum \cite{abrate2013slamming, truscott2014water, jung2021swimming}. One method for mitigating the impact loading is to modify the impactor geometry. For instance, a finely tapered geometry such as a narrow cone encounters a much lower peak impact force than a blunt geometry \cite{baldwin1971vertical}, consistent with the morphology of plunge diving sea birds \cite{chang2016seabirds}. Another approach to peak force reduction is to introduce elastic compliance to soften the impact. Previous experimental studies have primarily examined the response of continuously deformable flexible wedges as a model for a ship hull \cite{gilbert2023wedge}. Even in the case of a rigid impactor, when the tip is sufficiently blunt the trapped gas layer can be modeled as a piston cylinder, contributing an effective elasticity to the problem \cite{belden2024sphereflat}. Yet relatively few parametric studies with simplified flexible impactors have been performed \cite{wu2020water, boom2023water}, which is necessary towards developing a generalized understanding of the effects of flexibility. In our previous experiments with a simplified flexible impactor consisting of two masses separated by a linear spring, we directly measured the force of impact as a function of speed and stiffness \cite{antolik2023shm}. Consistent with the intuition for a cushioned impact, the experiments demonstrated that the impact force on the trailing body can be reduced compared to the rigid case when a sufficiently soft spring is used to connect the nose. Surprisingly, however, we also observed that the force of impact can increase dramatically compared to the rigid case if the structure's natural period of oscillation is small relative to the time scale of impact (inversely proportional to the impact speed). Thus a certain compliance may be beneficial at one speed but detrimental at others, suggesting that passive linear elastic compliance may not be a suitable force reduction technique for engineered structures that typically must operate over a wide range of impact speeds and conditions.

When studying a coupled fluid-structure interaction problem experimentally, a common approach is to use a so-called cyber-physical system in order to efficiently traverse the parameter space by programming the structural parameters into a computer-based force-feedback controller that enacts the desired physical behavior with a mechanical actuator, allowing structural couplings beyond the simple linear elastic case to be explored. The concept of the cyber-physical system for fluid mechanics experiments was pioneered by Miller and Hover \cite{miller1996cps, hover1997vivmarinecables, hover1998oscillatingcylinders} and typically takes the form of a large facility which includes the electronic control system and mechanical actuator attached to the test section of a water flume, tow tank, or wind tunnel. Cyber-physical systems have been used to experimentally study vortex induced vibrations of cylinders in a flow \cite{mackowski2011cps, lee2011viv} as well as the response of elastically mounted pitching airfoils \cite{onoue2016cpspitchingplate, zhu2020cpsflume, fagley2016aeroelastic}, among other applications. Recently the concept of the robotic test system has been extended to emphasize automated experimentation by Fan \textit{et al.} with the construction of a robotic tow tank that algorithmically traverses the parameter space and continuously performs experiments even while unattended \cite{fan2019towtank}. A conceptually similar experimental approach has been employed to study swimming in fish-like robots in which a software-controlled actuator is used to adjust the robot's tail stiffness on the fly \cite{zhong2021fish}. Likewise, decimeter-scale programmable stiffness spring elements which utilize voice coil actuators and strain gauge displacement sensors were designed by Lee \textit{et al.} to make up the connections between nodes of a mechanical neural network that could learn different bulk material behaviors \cite{lee2022mnn}. Despite the extensive use of cyber-physical systems and adjacent concepts in the literature, to our knowledge the CyberDiver is the first such fully \textit{untethered} system to be developed and applied to fluid-structure interaction experiments. We demonstrate that the CyberDiver can operate as a high-bandwidth position feedback cyber-physical system and accurately simulate the target structural coupling during water-entry experiments with typical time scales on the order of 10 ms. By implementing structural damping or nonlinear restoring force laws, the CyberDiver can more effectively mitigate impact loading than a linear stiffness flexible impactor, directly informing and motivating designs of more advanced passive structures for force mitigation.

Beyond passive flexibility, a number of examples in the biological world perform active maneuvers during water entry. Diving sea birds have been observed to tuck their wings in an active maneuver that is coordinated with the moment of impact \cite{lee1981plummeting}, likely to minimize their drag and reach greater depths in the pursuit of prey \cite{brierley2001gannets, sharker2019water}. Similarly, human Olympic divers sometimes perform a ``rip'' entry maneuver by bending at the waist shortly after entering the water to reduce their splash, an effect which has recently been experimentally replicated using a simplified diver model with a single passive hinge \cite{gregorio2023air}. In the Māori sport of Manu jumping, a different maneuver is performed in which competitors form a V-shape while diving and rapidly expand outward shortly after entering the water, this time with the intent of creating the largest splash. Rohilla \textit{et al.} replicated this effect with a simplified spring loaded wedge shaped mechanical diver that opens shortly after impact \cite{rohilla2024manu}. By operating the CyberDiver in a position control mode and actively extending or retracting the nose in a maneuver synchronized with impact, the splash crown formation can be amplified or suppressed, demonstrating that the CyberDiver is a viable framework to perform tightly controlled parametric experiments on the influence of active motion during water entry, a problem which has not yet received much attention.  

\section*{Results}
\subsection*{Biological divers inspire active system for water-entry experiments}

Inspired by the examples of sea birds and human divers, but occupying a better defined parameter space, we designed a slender axisymmetric impactor with a single actuated degree of freedom in which the lightweight hemispherical nose moves axially relative to the heavier trailing body. In our experiments, the impactor is dropped from various heights into a large water bath as shown in Figure \ref{fig:fig1}(A) and Movie S1, forming a splash as it enters the water and experiences hydrodynamic loading, potentially producing a trailing air cavity as it plunges deeper into the water. An exploded view of the computer-aided-design (CAD) model of the CyberDiver is shown in Figure \ref{fig:fig1}(E). The nose is attached to the body via a pair of spring steel disk flexures that allow axial motion with very repeatable restoring force and no static friction while exhibiting high stiffness in the other degrees of freedom. A highly compact optical linear encoder mounted between the disk flexure bearings with 1 $\mu$m resolution and speed up to 14 m/s (Celera CE300-40) is used to measure the instantaneous nose displacement and implement feedback control. A voice coil (BEI Kimco LA15-16-020A) is used as the mechanical actuator due to its ability to apply force with high slew rate in the axial direction while similarly producing no sliding friction that would affect the impact dynamics. The overall impactor diameter is minimized in order to balance the impact loading (which scales with the diameter squared) and the peak force of the voice coil actuator. Maintaining an overall diameter that does not significantly exceed the voice coil diameter motivated housing the control electronics and battery in the slender trailing body. An accelerometer array on the trailing body of the impactor records its deceleration during impact, a quantity that is to be minimized to avoid damage to sensitive payloads in applications. 

By using the instantaneous displacement and velocity feedback from the linear encoder and applying a proportional restoring force with the voice coil actuator, the CyberDiver can behave as a position feedback cyber-physical system emulating a passively flexible damped harmonic oscillator, for instance, as shown in Figure \ref{fig:fig1}(B). Both the stiffness and the damping ratio of the structural coupling can be programmed to a large range of values, limited only by control loop stability and the peak force that the voice coil can produce. Typical results for the acceleration of the trailing body after impact are shown in the inset in Figure \ref{fig:fig1}(B) for impacts at 4 m/s with 15 N/mm programmed linear stiffness and cases with zero and intermediate prescribed structural damping (damping ratios $\zeta=0$ and 0.25). Contrary to the highly impulsive hydrodynamic loading experienced by a rigid impactor, the cyber-physical spring-damper buffer extends the time scale of the slamming phase forcing and in this case reduces the peak deceleration experienced by the trailing body. The oscillations continue for several periods in the $\zeta=0$ case as they slowly decay due to the effects of minor material damping and fluid friction.

After comparing the performance of the active cyber-physical system to the passive linear case and finding good agreement, we use the CyberDiver to explore new nonlinear structural couplings that can produce even greater reductions in the impact loading experienced by the trailing body. In particular we explore the family of nonlinear curves illustrated in Figure \ref{fig:fig1}(C) with a linear region near the origin and a programmable force limit at which the restoring force saturates, producing a zero stiffness region. In the impact experiments with this nonlinear structural coupling, any peaks in the acceleration profile that would otherwise exceed the programmed force limit are mitigated, as shown in the corresponding accelerometer recording. Notably, this technique works regardless of impact speed as long as there is sufficient nose travel, and thus overcomes a major limitation of the linear structural coupling which only effectively reduces the impact loading at particular speeds.

Finally, the CyberDiver can operate in a control mode where the nose displacement is prescribed in time relative to the moment of impact as shown in Figure \ref{fig:fig1}(D) or Movie S2. Thus we can approximate the active maneuvers of biological systems like human competitive divers or sea birds in experiments and modify impact quantities such as the splash size in a repeatable fashion within a simplified parameter space. We demonstrate that extending the impactor nose at the moment of impact significantly amplifies the size of the splash crown whereas retracting the nose at the moment of impact reduces its size.

\subsection*{Mechanical and electrical subsystems designed for high performance on fast impact timescales}

The isolated motion subassembly of the CyberDiver is shown in Figure \ref{fig:fig2}(A) with laser cut spring steel disk flexure bearings connected to the hemispherical nose through a central shaft. The coil component of the actuator is attached to the end of the central shaft and its magnetic housing is attached to the impactor body. The encoder's glass scale is fastened to the central shaft between the disk flexures and opposite the read head PCB mounted to the impactor body. A complete description of the dynamics of the CyberDiver includes the electronics as well as the mechanical motion. The simplified electric circuit for the voice coil actuator is shown in Figure \ref{fig:fig2}(A) in which the voice coil actuator is modeled as a series resistor with resistance $r$ and inductor with inductance $l$. The power electronics apply a voltage $v$ across the coil and measure the resulting current using a sense resistor with negligible resistance compared to the coil. The voltage in the coil is coupled to the mechanical displacement of the nose through the back electromotive force (EMF) since relative motion compared to the magnetic housing induces a potential difference in the coil \cite{desilva2015sensors}. The magnitude of the back EMF can be approximated as $k_b(\delta) \dot{\delta}$ where $k_b(\delta)$ is the back EMF constant and the nose displacement $\delta = x_b - x_n$ is the difference between the positions of the impactor body and nose. In practice $k_b$ is nearly constant but does change somewhat depending on the coil's position in its axial travel range. Combining the contributions of the elements in the circuit, the equation for the coil current $i$ is

\begin{equation} \label{eq:coil}
    v - k_b(\delta) \dot{\delta} = l \frac{\textrm{d}i}{\textrm{d}t} + ir.
\end{equation}

Equation \ref{eq:coil} demonstrates the need for closed loop control because back EMF will act as a disturbance that influences the coil force. If left unchecked, the back EMF tends to subdue the motion of the actuator and hence would be unsuitable for emulating an undamped spring structure. Additionally equation \ref{eq:coil} shows that even in the absence of back EMF the coil current will lag behind changes in the applied voltage by approximately $l/r = 318 \textrm{ $\mu$s}$ for our coil, which is much faster than the time scale of impact ranging from 4.5 to 13.5 ms, but can be improved further with closed loop control. To implement the cyber-physical system or perform active maneuvers at impact we wish to specify a voltage $v$ to rapidly and accurately control the force produced by the coil, which is given by $k_f(\delta) i$ where $k_f(\delta)$ is the coil force coefficient, again varying throughout the travel range of the voice coil. 

The coil dynamics are coupled to the mechanical dynamics in the equations of motion for the impactor nose and body. As shown in Figure \ref{fig:fig2}(B), the flexure outer ring is rigidly clamped to the impactor body and the inner ring is clamped to the central shaft. The cutouts in the flexures are designed targeting a linear response in the axial direction and high stiffness in other directions. In addition to the forces applied by the actuator and the hydrodynamic loading during impact, the nose also experiences a passive restoring force from the disk flexure bearings. The net nose mass is comprised of the mass of the hemisphere, the central shaft, the coil, the sealing shell, and half the mass of the disk flexures. The rest of the mass in the CyberDiver is rigidly affixed to the trailing body. Hence the equation of motion for the impactor body is
\begin{equation} 
    (1-\alpha)M\ddot{x}_b = -F_p(\delta) + k_f(\delta) i + F_b(t) 
\end{equation}
and the equation of motion for the impactor nose is
\begin{equation}
    \alpha M \ddot{x}_n=F_p(\delta) - k_f(\delta) i + F_n(t)
\end{equation}
where $M$ is the overall mass of the CyberDiver, $\alpha$ is the ratio of the nose mass to the total mass, $F_p(\delta)$ is the passive restoring force of the disk flexure bearings (which we approximate in practice as a weakly cubic spring) and $F_b$ and $F_n$ are external forces. In our experiments $F_b = 0$, as the hydrodynamic loading is isolated to the nose.

Figure \ref{fig:fig2}(D) illustrates the control scheme used to operate as a cyber-physical system and prescribe the structural dynamics of this tightly coupled electromechanical system. Responsibilities are divided between the power board and the control board as shown in Figure \ref{fig:fig2}(C). The control board calculates a target coil force which takes into account both the configured structural coupling $F_s(\delta)$ and calibration curves for the passive restoring force $F_p(\delta)$ and coil force response $k_f(\delta)$. The set point coil current is calculated by combining these contributions as 
\begin{equation}
    i_\textrm{sp} = \frac{F_p(\delta) - F_s(\delta) - c \dot{\delta}}{k_f(\delta)}
\end{equation}
where $c$ represents the configured damping coefficient and $\delta$ is the displacement measured by the encoder. In the damping force calculation, $\dot{\delta}$ is estimated using the Savitsky-Golay method \cite{schafer2011savitzky} with 7 sample frames and second-order polynomials. The power board receives the command for the target coil current and implements a PID controller to regulate the coil current. The PID controller includes a feed forward term to better respond to a rapidly changing set point without impacting the controller stability. The control loops on both the control board and power board update at a frequency of 50 kHz. 

We developed and implemented custom electronics to accurately control the coil force during impact, perform the real time control calculations, and log experimental data as shown in Figure \ref{fig:fig2}(C). The system is powered from an off-the-shelf lithium polymer battery, enabling fully untethered operation of the CyberDiver as it plunges into the water bath. Rather than using a single-chip H-bridge, the power board features an H-bridge designed from discrete components in order to support the large 5.7 A maximum current required to drive the coil, as well as a dedicated microcontroller (STM32F405) to run the coil current PID loop. The control board uses a separate microcontroller (STM32H723) to calculate the target coil current depending on the operating mode and communicate this value to the power board over a Serial Peripheral Interface (SPI) data link. Additionally, the control board acquires and logs data to a Secure Digital (SD) card from an array of high performance micro-electromechanical system (MEMS) accelerometers and from the encoder daughter board.

Figure \ref{fig:fig2}(E) shows the response of the voice coil current to a 0.5 A step change in the commanded current. For these tests the impactor is suspended from a low-stiffness bungee with the intent of capturing the dynamics of the free system. The open-loop step response reveals complicated dynamics due to the interplay between the coil forcing and the back EMF from the resultant nose displacement velocity, stabilizing on timescales that are substantially longer than impact. On the other hand, the closed loop controller with aggressively tuned gains stabilizes at the set point value within approximately 500 $\mu$s with an acceptable amount of overshoot. This test in fact represents a worst case scenario because the target coil current tends to change more gradually when subjecting the cyber-physical system to impact loading since the programmed force-versus-displacement curves tested herein are continuous. 

\subsection*{Untethered cyber-physical system accurately simulates linear stiffness structure during water entry}

To demonstrate its viability as a cyber-physical system, the CyberDiver is programmed to operate as a linear spring-mass system with adjustable stiffness. Both static and dynamic physical testing are performed to confirm its behavior. The results of the static testing are shown in Figure \ref{fig:fig3}(A) in which the robotic diver is compressed axially at a rate of 0.25 mm/s in a custom force testing machine. The restoring force of the cyber-physical structural coupling is measured using a load cell and the response is plotted for the programmed stiffness values of 5, 15, or 45 N/mm. For the case of 45 N/mm programmed stiffness, we also introduce a slight programmed damping coefficient of 0.005 N/(mm/s) (corresponding to a damping ratio $\zeta = 0.03$) to stabilize the controller and prevent the slight vibrations that can occur due to feedback overshoot. The other cases of 5 and 15 N/mm programmed stiffness do not include an artificial damping term. In all cases the measured force agrees well with the target structural curve indicating that the feedback controller and voice coil actuator can accurately produce a target force. The extents of each compression test are defined such that they respect both the 6 mm compression limit of the diver and the 33.8 N peak force that the actuator can produce.

We also measure the impulse response of the CyberDiver to confirm that it matches the expected analytical result for a system of two masses connected by a spring. The CyberDiver is suspended from a low-stiffness bungee and struck with an impact hammer while executing the three simulated stiffness cases. The displacement response is recorded with the onboard linear encoder and plotted on normalized axes in Figure \ref{fig:fig3}(B), and also visualized in Movie S3. The impactor's rigid body mode while hanging from the bungee has frequency 1.2 Hz while the expected frequencies of the impactor vibration range from 32 Hz for the 5 N/mm case to 96 Hz for the 45 N/mm case. Thus, due to the large frequency separation from the rigid body mode, the measurements are a faithful representation of the response of the free system. The plot's ordinate is normalized by the maximum displacement during the test so that all trials are clearly visible, since the dimensional displacement is considerably lower for cases with higher stiffness. The abscissa is normalized by the expected natural frequency of the free two mass system $f_n = 1/(2\pi) \sqrt{k / (\alpha (1-\alpha) M)}$ such that we expect one period of oscillation to occur for each horizontal unit of the plot. Here $k$ is the programmed stiffness of the virtual spring, $M = \textrm{ 0.792 kg}$ is the total mass of the impactor, and $\alpha = 0.193$ is the ratio of the nose mass to the total mass. These mass values differ slightly in the water impact experiments because the transparent sealing shell is removed during impulse response measurements in order to attach the bungee. The actual oscillation period of the cyber-physical system closely matches the expected analytical value as the measured oscillations remain synchronized with the normalized scale for several periods in Figure \ref{fig:fig3}(B). Additionally, an envelope of oscillation decay rates corresponding to values of the damping ratio $\zeta$ between 0.015 and 0.045 is overlaid on the experimental measurements indicating that the system experiences very little damping for the three programmed stiffness values, and thus closely approximates the behavior of the desired undamped system. The very weak damping measured in practice likely results from material damping in the flexure bearings and latency in the controller's efforts to reject back EMF.
 
With the static and dynamic behavior of the programmed structural dynamics validated, we performed water entry experiments in which the CyberDiver plunges into a large quiescent water bath with normal incidence at speeds $V$ ranging from 2 to 6 m/s. The overall geometry of the impactor when fully assembled with the transparent outer sealing shell is a slender cylinder with a hemispherical nose whose radius $R$ is 27 mm. The overall length of the impactor is 256 mm. Because the shell of the CyberDiver is rigidly fixed to the hemispherical nose, the geometry of the impactor surface in contact with the water does not change throughout an impact; rather, the impact dynamics are only influenced by momentum transfer between the nose and body masses mediated by the active control. When the CyberDiver is fully assembled with the shell, the total mass $M$ is 0.828 kg and the ratio of the nose mass to the total mass $\alpha$ is 0.228. The water's density is represented by $\rho$, its viscosity by $\mu$, and the surface tension of the air-water interface by $\gamma$. The Reynolds number $\rho V R / \mu$ in our impact experiments ranges from $5.4 \times 10^4$ to $1.6 \times 10^5$, the Weber number $\rho V^2 R / \gamma$ from $1.5 \times 10^3$ to $1.4 \times 10^4$, and the Froude number $V / \sqrt{gR}$ from 3.9 to 12. As a result we expect the hydrodynamic forcing during water entry of our impactor to be dominated by fluid inertia, with significantly smaller contributions from viscosity, surface tension, and hydrostatics. When the free falling impactor collides with the water surface, it experiences an impulsive load from the sudden hydrodynamic resistance. This impulsive loading excites the cyber-physical structural mode of the CyberDiver as shown in Figure \ref{fig:fig3}(C) for impacts at 4 m/s. As it plunges deeper into the water, the structure continues to oscillate but the amplitude decays over time due to structural damping and interactions with the fluid. As such the peak deceleration of the water entry event tends to occur at early times and is indicated in the experimental data here with a square marker. The shaded regions indicate the standard deviation between five experimental replicates. The experimental data is captured using the IIS3DWB accelerometer chip sampling at 26.6 kHz with 1.3 kHz bandwidth. The accelerometer is soldered to the control PCB so the measurement captures the acceleration of the trailing body of the impactor which might contain the sensitive payload in engineering applications. The rigid theory curve represents the acceleration that would be felt by a fully rigid spherical impactor having the same radius and overall mass as the CyberDiver. The theory is due to Shiffman and Spencer \cite{shiffman1945sphere1, shiffman1945sphere2, shiffman1947lens} and has been widely used and experimentally validated in the literature \cite{moghisi1981experiments}, including in our prior work \cite{antolik2023shm}. Notably, all of the linear spring impact experiments here spread out the time scale of the hydrodynamic slamming as compared to the rigid case, consistent with the typical intuition for an impact cushioned by a spring. For the 5 N/mm and 15 N/mm springs, the peak deceleration is notably lower than in the rigid case. However, in the stiffest case with 45 N/mm, the peak deceleration is higher than the rigid case. The CyberDiver replicates this finding documented in our prior work with a passive flexible impactor with linear stiffness, and highlights a primary limitation of using a linear structural element to curtail slamming forces during water entry. The transition between force reduction and amplification as the stiffness of the structural coupling increases is captured in a dimensionless parameter consisting of the ratio between the time scale of the hydrodynamic loading and the natural period of oscillation of the impactor. This parameter, called the hydroelastic number, is defined as
\begin{equation}
    R_F = \sqrt{\frac{k}{\alpha (1-\alpha) M}} \frac{R}{V}.
\end{equation}
When the natural frequency of the impactor $\sqrt{k/(\alpha (1-\alpha) M)}$ is sufficiently large compared to the hydrodynamic time scale $R/V$ (this is the time that it takes for the impactor to reach one radius depth assuming minimal slowdown), the structural coupling begins to oscillate while the hydrodynamic forcing is ongoing and this combined forcing produces a larger net deceleration in a resonance-like phenomenon \cite{antolik2023shm}.  

For a fixed nose geometry and mass distribution, the hydroelastic number uniquely dictates whether the peak impact loading will increase or decrease as compared to the rigid impactor, as shown in Figure \ref{fig:fig3}(D). This figure presents experiments conducted with the CyberDiver at many combinations of linear stiffnesses and impact speeds, and the normalized peak force is plotted against the hydroelastic number. Here, the peak acceleration on the trailing body is multiplied by the total impactor mass and divided by an inertial fluid force scale to produce a peak impact drag coefficient for each experimental case. The horizontal line at 1.05 is the theoretical peak impact drag coefficient for a rigid sphere. Thus the criterion for a flexible linear spring impactor (with the chosen geometry and mass distribution) that effectively reduces the impact loading is $R_F \lessapprox 2.5$; otherwise the addition of the spring is detrimental. In particular, $R_F$ tends to be large for cases with high stiffness or low entry speed. The added mass model curve that was derived in our prior work \cite{antolik2023shm} and is detailed here in the Supplementary Text agrees well with the experimental results. In applications, the designer of the structural coupling can control the stiffness but it may not be possible to control the entry speed -- in future sections we explore different structural couplings enabled by the cyber-physical system that can overcome this limitation and successfully mitigate impact loading regardless of entry speed. 

Note that there are several acceleration features throughout the full impact -- the initial slamming phase, development of steady state drag, cavity formation and pinch off, and ultimately collision with the bottom of tank. We only report the first peak here, associated with the slamming phase. Sometimes, at low speeds or low stiffness, the steady drag can develop to be a larger value than the slamming peak. Additionally, hitting the bottom of the tank is always the most violent event in our experiments. In applications, depending on the impact conditions or the particular impactor geometry, these additional acceleration events may need to be considered as well.

\subsection*{Structural damping reduces peak impact loading}

In a canonical oscillating system, the introduction of damping reduces the vibration amplitude when excitation is applied near the system's natural frequency. Since the transition between force reduction and amplification in the flexible impactor system is reflected in a ratio of timescales reminiscent of the classic resonance effect, it is reasonable to expect that impact loading could be reduced by introducing damping in the structure. To explore this idea we performed a set of experiments in which the CyberDiver was programmed with damping ratios $\zeta$ of 0.25 and 0.5 and compared against the results from the nominally undamped case (whose damping ratio is approximately 0.03). The ability of the cyber-physical system to replicate the behavior of a damped harmonic oscillator was tested by performing impulse response measurements using the same methods as the undamped case, and the results are shown in Figure \ref{fig:fig4}(A). For all of the damped experiments the programmed spring element of the structure has a linear stiffness of 15 N/mm. The vibrations of the damped impactor subside quickly compared to the undamped case and the decay rate closely matches the theoretical exponential envelope curves for the given programmed damping ratio. Hence the CyberDiver can accurately behave as a damped harmonic oscillator with programmable stiffness and damping. Due to stability limitations with the controller, critical damping $\zeta = 1$ could not be achieved with the present system for the tested parameters.  

The deceleration of the damped impactor as it enters the water at 3 m/s is shown in Figure \ref{fig:fig4}(B) for two different damping ratios and compared to the undamped case. The peak deceleration in each case is again indicated with a square marker. The undamped case corresponds to a hydroelastic number that exceeds the critical value and thus experiences greater peak deceleration than the rigid case. However, by introducing damping, the peak deceleration is reduced to values below the rigid peak. The measured peak impact drag coefficient as a function of hydroelastic number is reported in Figure \ref{fig:fig4}(C) where the stiffness is fixed at 15 N/mm and several combinations of speed and damping ratios are tested. For this range, the $\zeta=0.5$ case performs the best and reduces the peak flexible impact drag coefficient below or equal to that of the equivalent rigid case for all speeds tested. The benefits of damping are most pronounced at high $R_F$ values. The experimental results agree well with the added mass model when solved with the programmed damping values, indicating that the simple model has predictive capabilities that extend beyond the case of the undamped simple harmonic oscillator validated in our prior work. In summary, both the experiments and model suggest that the inclusion of damping (in the underdamped $\zeta<1$ regime) can serve to increase the critical hydroelastic number, thereby broadening the range of speeds where introducing flexibility is beneficial to peak load mitigation. Nevertheless, damping is not a one-size-fits-all solution to the problem, as the impact force can still exceed the rigid case provided the hydroelastic number is sufficiently high. Additionally, at lower hydroelastic numbers than shown here, the theory predicts that the undamped case will experience less loading than the damped case, a trade-off consistent with the physics of passive vibration isolation.

\subsection*{Nonlinear structural coupling clamps peak acceleration to a programmable limit} \label{sec:cyberdiver_nonlinear_results}

In engineering applications, a common constraint is that the sensitive payload in the trailing body may suffer damage if it experiences decelerations above a certain threshold. As we have seen thus far, a flexible impactor with a linear spring is not guaranteed to mitigate impact deceleration compared to a rigid impactor when it operates over a wide range of entry speeds. Furthermore, the dimensional peak deceleration still increases monotonically with impact speed, albeit at a slower rate than the rigid case. A more robust approach would be to implement a structural coupling that limits the peak deceleration directly. As illustrated in the schematic in Figure \ref{fig:fig5}(B), the trailing body only experiences force (and thus acceleration) through the prescribed structural coupling. Therefore, if we wish to impose a strict limit on the acceleration felt by the trailing body, we can implement a nonlinear structural coupling as shown in Figure \ref{fig:fig5}(A). Near the origin, the structure behaves as a linear spring with, in this case, 15 N/mm stiffness. However, the force curve levels off at some programmable threshold away from the origin and behaves in this region as a zero-stiffness structure. The maximum acceleration that this structure can apply to the trailing body is simply the force threshold divided by the mass of the trailing body, allowing a deceleration limit to be directly imposed. The advantage of this technique is that the controller requires no \textit{a priori} knowledge of the impact parameters and therefore the coupling could be implemented passively with an appropriately designed quasi-zero stiffness (QZS) nonlinear structure \cite{yan2022qzs, chai2024qzs}. Figure \ref{fig:fig5}(A) shows the measured force-versus-displacement curve of the CyberDiver as it is compressed in the force test machine while programmed with the nonlinear structural law with force thresholds ranging from 5 N to 20 N. The actual behavior closely matches the overlaid target threshold lines demonstrating that the CyberDiver can effectively behave as a zero stiffness structure. 

In our drop tests, the active control performs the acceleration limiting successfully on the rapid time scales of impact, and effectively clips the deceleration peaks that would otherwise exceed the programmed force threshold. Figure \ref{fig:fig5}(B) shows the deceleration during impacts at 5 m/s for cases with 15 N/mm linear stiffness, a nonlinear curve with 10 N force threshold, and one with 15 N threshold. As the acceleration peaks are clipped, the period of oscillation of the impactor tends to lengthen, consistent with the physics of a softening spring. Figure \ref{fig:fig5}(C) reports the peak deceleration versus impact speed for the case of the 15 N/mm linear spring as well as each of the four nonlinear curves tested. Since the linear region of the nonlinear curve near the origin has 15 N/mm stiffness in all cases tested, all of the experimental curves in Figure \ref{fig:fig5}(C) converge at low speeds where the hydrodynamic loading is not sufficiently strong to push the structure out of the linear region. At higher speeds however, the curves diverge as each case levels off at its respective target acceleration limit. The known theoretical curve for the rigid case with its quadratic scaling is shown for reference. As expected given the existence of a critical hydroelastic number as previously discussed, the 15 N/mm linear case crosses the rigid curve at some intermediate speed and, for high speed, attenuates the peak deceleration significantly compared to the rigid case. However, the peak dimensional deceleration of this linear stiffness case still increases dramatically with speed, albeit at slower than the quadratic rate of the rigid case. The nonlinear structural curves perform significantly better across the full range of impact speeds and can be directly tailored based on the requirements of the impactor payload.  

A notable limitation of this method of mitigating impact loading is the finite travel of the joint between the impactor nose and body, which enters in as an additional design constraint. In the nonlinear impact cases with lower force threshold, the nose is compressed to a greater extent. If the hydrodynamic loading is too high and the force threshold programmed too low, the impactor nose will simply bottom out against the trailing body at some point during the during of the water entry, resulting in a hard collision and large accelerations. The experimental cases where the impactor bottoms out before colliding with the bottom of the tank are circled in Figure \ref{fig:fig5}(C). For these points the peak deceleration before bottoming out is reported. In the other cases the travel limit is not exceeded in the time before the impactor collides with the bottom of the tank. In the Supplementary Text we demonstrate that increasing the stiffness in the linear region can extend the operating envelope in which the impactor does not bottom out. The strategy presented here requires no \textit{a priori} knowledge of the impact speed or when the impact will occur, so it could be applicable to a wide range of situations.

\subsection*{Active maneuvers coordinated with impact influence splash size}

A number of examples of active maneuvering during impact occur in the biological world, including plunge diving sea birds that tuck their wings before impact \cite{lee1981plummeting} and, with particular relevance to the present study, human sport divers who reconfigure their shape during impact to either minimize or accentuate splash formation \cite{gregorio2023air, rohilla2024manu}. To perform active maneuvers during impact, the CyberDiver operates in a closed loop position control mode in which the axial nose displacement may be specified as a function of time relative to the moment of impact. A light emitting diode (LED) on the control board is turned on at the same time as the maneuver is initiated so that the timing can be verified in the high-speed video. Contrary to biological divers, the CyberDiver geometry does not change during the maneuver -- rather an acceleration can be applied to the impactor nose, presenting an opportunity to explore splash manipulation techniques with a simplified model of an active impactor. In our splash manipulation experiments, the impact speed is 2 m/s (constant drop height) and a super-hydrophobic coating (Rust-Oleum NeverWet) is applied to the anodized aluminum nose of the CyberDiver so that the ejecta sheet separates during splash formation and a highly visible splash crown is produced. In the active maneuver cases, the set point of the position controller is changed at the moment of impact and the nose is either retracted, reducing the size of the splash, or extended, increasing the size of the splash. Photographs of the splash 75 ms after the moment of initial impact are shown in Figure \ref{fig:fig6} and in the corresponding Movie S4, illustrating a dramatic change in the size of the splash depending on the maneuver performed. The photographs show the splashes at approximately their largest sizes throughout the impact. In cases B and D, the nose displacement during free fall is zero and then the nose is commanded to retract or extend by 4 mm at the moment of impact. In cases A and E, the nose is already displaced by $\mp$4 mm during free fall so that the magnitude of the position change during the maneuver is greater, producing an even larger change to the splash size. In each of the five splash cases, three experimental replicates are performed to ensure that the modification to the splash size is repeatable. Figure \ref{fig:fig6}(F) reports the commanded and actual displacement of the nose from one trial of each of the maneuver cases. The CyberDiver reaches the target position approximately 10 ms after the command is issued, with some settling time thereafter due to the controller overshoot. The response time of position control loop is slower than the current control loop due to inertia. In retraction case A, the nose bottoms out at 6 mm as it overshoots the commanded displacement, helped along by the developing hydrodynamic force, but no damage to the CyberDiver is sustained. In the control case C with no maneuver, the displacement remains within 0.2 mm of the target neutral position in spite of the disturbance due to the hydrodynamic forcing. These results demonstrate that the CyberDiver is a viable framework for further experiments with active maneuvering during impact, in which the axial displacement profile and timing can be tightly controlled. A more detailed series of experiments in the future investigating the dependence of the splash size on both the maneuver amplitude and its timing, as well as the impact speed, could have implications for better understanding sport diving and biological divers in general.        

\section*{Discussion}

We have presented the CyberDiver, a compact untethered robotic impactor that can be programmed to behave as a cyber-physical system with arbitrary passive flexibility or to perform active maneuvers. As it plunges into the water, the CyberDiver simultaneously records high-speed measurements of the dynamics, enabling highly resolved parametric experimental studies of flexible water entry. Through static force measurements and dynamic vibration testing, we demonstrate that the CyberDiver can accurately produce target structural dynamics with natural frequencies up to nearly 100 Hz. A set of validation experiments using a linear structural coupling with programmable stiffness between the impactor nose and body to buffer the impact forcing replicates prior results with a similar passive flexible impactor and recalls the limitation that a linear spring is only beneficial in a narrow range of impact parameters. At low speeds or high stiffness, corresponding to high hydroelastic number $R_F$, the introduction of elasticity can actually be detrimental and produce greater impact loading on the trailing body than in the case of a rigid impactor, due to a resonance-like phenomenon. Utilizing the CyberDiver as a cyber-physical system, we demonstrate that the addition of damping to the linear structural coupling can substantially reduce the peak impact forces at moderate to high $R_F$, and increase the range over which compliance is beneficial, a finding supported by our two-way coupled added mass model for impact. However, robustly limiting the maximum force on the trailing body requires a different solution. To this end, we test a nonlinear structural coupling which has a linear region near the origin but plateaus to a constant force (zero stiffness) region away from the origin and show that the peak acceleration felt by the impactor body can be directly specified by programming the magnitude of the force-plateau region. Although an active system may not be feasible in many applications, there exist designs in the literature for passive quasi-zero stiffness (QZS) structures whose response closely resembles our programmed nonlinear structure \cite{chai2024qzs, yan2022qzs}. Thus we show that the CyberDiver may serve as a powerful design tool for engineered impacting structures, permitting a tailored structural coupling that is designed in software to be physically tested and exposed to the complete set of rich multi-scale fluid physics that it would encounter in a real application. Finally, we explore the effects of active maneuvering during impact on the size of the splash formed, inspired by biological examples such as sea birds and human sport divers \cite{gregorio2023air, rohilla2024manu, lee1981plummeting}. By operating the CyberDiver in a position control mode that synchronizes a nose extension or retraction with the moment of impact and observing the change in splash size, we lay the groundwork for a systematic parametric exploration of the effects of active motion on impact dynamics. As our understanding of the biological divers grows, we will in turn be increasingly able to apply elements of their design to engineered impactors.

\section*{Materials and Methods}
\subsection*{Experimental setup} \label{sec:expt_setup}

In order to measure its performance during water entry, the CyberDiver is dropped with normal incidence onto the surface of a quiescent water bath at speeds $V$ ranging from 2 to 6 m/s while the active control runs in various configurations. For repeatable entry speeds, the CyberDiver is suspended from an electromagnet above the tank mounted to an arm with adjustable height. The height of the electromagnet is set using a plumb line with markings such that the distance between the water surface and tip of the impactor nose in the neutral position is $H = V^2 / (2g)$. A steel ball is bolted to the end of the CyberDiver housing to form a point contact with the electromagnet that ensures the CyberDiver hangs vertically before a drop. The water bath cross-section is a square whose length and width are 76 cm with the intent of approximating an infinite domain compared to the impactor with 27 mm radius. The depth of the bath is 70 cm. A trampoline with an aluminum extrusion frame and rubber foam padding is placed at the bottom of the tank to avoid damage to the tank or the impactor during high-speed drops. The impacts are filmed at 20,000 frames per second from the side with a Phantom Veo 1310L camera equipped with a 55 mm Nikon lens at the height of the air-water interface in order to verify the speed of entry and observe the size of the splash formed during impact. 

\subsection*{Experimental procedure} \label{sec:expt_procedure}

Before running an impact experiment, the CyberDiver battery voltage is measured and, if necessary, an off-the-shelf lithium polymer battery charger is used. The CyberDiver is programmed to enter an error state and alert the user when the battery voltage drops below 23 V to ensure a repeatable voltage level in experiments. The drop arm is set to an appropriate height based on the desired entry speed and the binary configuration file for a given experiment with the CyberDiver is generated using a MATLAB script. All of the parameters of the active control for a given experiment are encoded in the binary configuration file that the CyberDiver reads into memory from an SD card at startup. The file includes a description string for the experiment, timing and triggering parameters for beginning the experiment, calibration curves, logging parameters, accelerometer configuration, control loop gains, cyber-physical parameters, and the experimental maneuver sequence. After the CyberDiver loads the configuration, the absolute position of the encoder must be found by manually compressing or extending the nose until the encoder index passes by the read head. The experiment is then initialized by pressing a physical button on the control board. The CyberDiver remains idle for the configured staging time (one minute in our experiments) while it is mounted to the electromagnet and any residual swinging motion is allowed to decay. The CyberDiver then waits until it detects free fall by checking if the absolute value of the accelerometer reading drops below 0.5 g. At this point data logging begins and the active control starts to run the configured experimental sequence. In our drop experiments, the CyberDiver remains in the idle mode for 50 ms after entering free fall to gain sufficient clearance to the electromagnet. It then enters the position control mode with a set point of zero displacement in order to remove any oscillations that originated from any nose displacement due to the effect of gravity while hanging from the magnet. Position control is used only briefly during free fall rather than for the relatively long duration that the impactor is suspended from the electromagnet to avoid excessive battery drain and coil heating. In the cyber-physical drop experiments, the CyberDiver switches to simulated structure mode 125 ms after entering free fall, which is well before impact with the water surface for the range of speeds tested. In the active maneuvering experiments, the CyberDiver remains in position control mode for the entire impact with the set point sequence shown in Figure \ref{fig:fig6}(F). Additionally, in the active maneuvering experiments, an LED on the control board is illuminated at the same moment as the active maneuver is initiated to verify its timing from the high-speed video. During the experiment, the CyberDiver logs data at 50 kHz consisting of microsecond time stamps, coil current measurements, calculated coupling force, encoder position and velocity measurements, coil H-bridge duty cycle, three axes of acceleration, operating mode, and the set point of the active feedback control loop.


\begin{figure} 
	\centering
	\includegraphics[scale = 0.65]{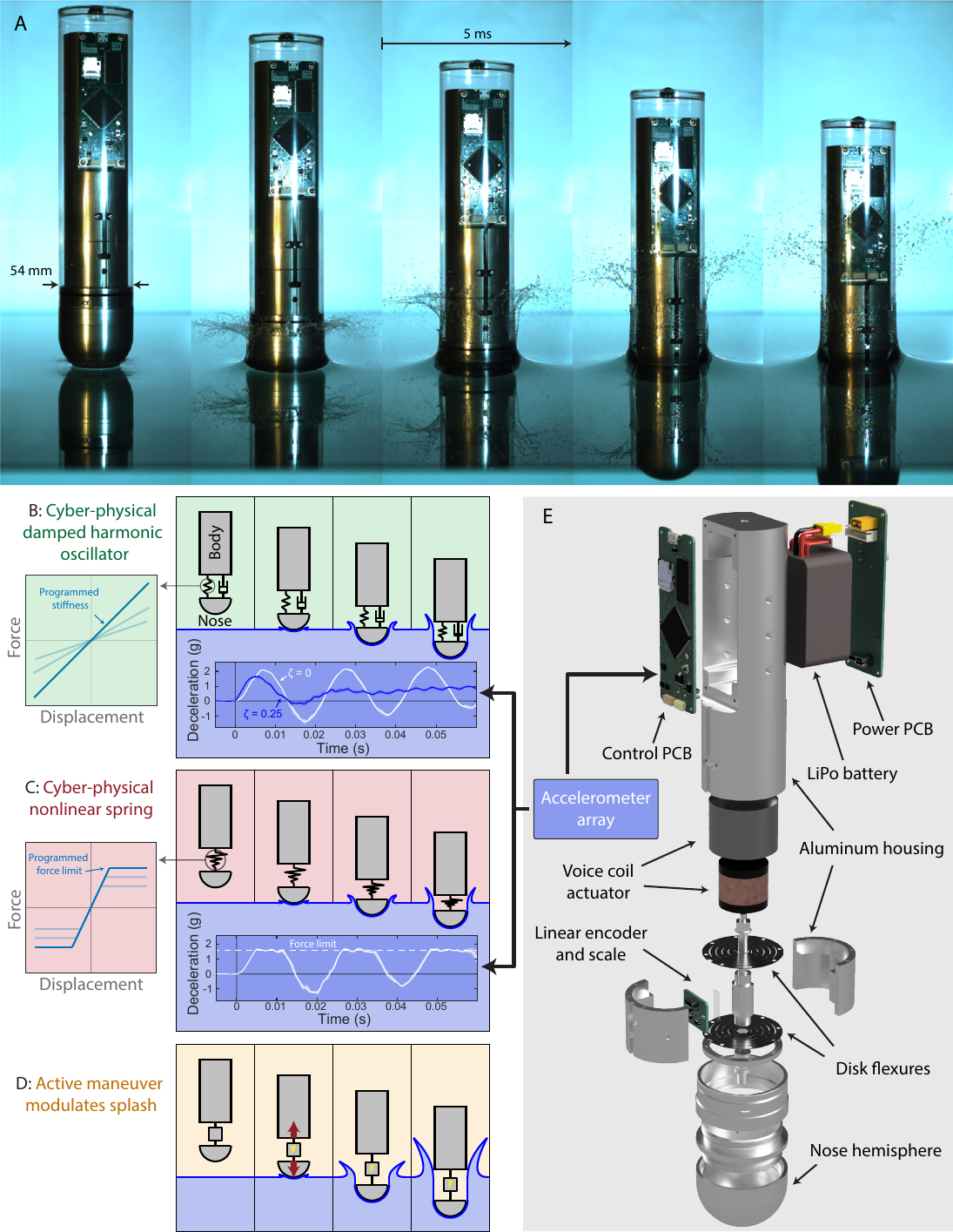}
        \caption{\setlength{\baselineskip}{12pt} \textbf{A robotic diver was fabricated that can perform active maneuvers during water entry.} (\textbf{A}) High-speed photographs of the CyberDiver as it enters a quiescent water bath with normal incidence at 4 m/s while simulating a linear spring with stiffness 5 N/mm. The time between frames is 5 ms. (\textbf{B}) The CyberDiver can be programmed to operate as a linear damped harmonic oscillator with programmable stiffness and damping coefficient. The inset plot shows the measured deceleration of the impactor body from drops at 4 m/s with 15 N/mm programmed stiffness and cases with zero or intermediate programmed damping. (\textbf{C}) The CyberDiver can also be programmed to behave as a nonlinear spring. Here the force-versus-displacement curve features a linear region near the origin and a threshold beyond which the force saturates, hence limiting the peak acceleration. The inset plot shows the measured deceleration of the impactor body from a drop at 4 m/s with 15 N/mm programmed stiffness in the linear region and a force limit of 10 N. (\textbf{D}) The CyberDiver performs active maneuvers that are synchronized with the moment of impact, modulating the size of the splash crown. (\textbf{E}) An exploded view of the CyberDiver reveals the actuator, sensing elements, control electronics, and mechanical disk flexure bearings. The sealing shell is not shown.}
	\label{fig:fig1}
\end{figure}

\begin{figure} 
	\centering
	\includegraphics[scale = 0.85]{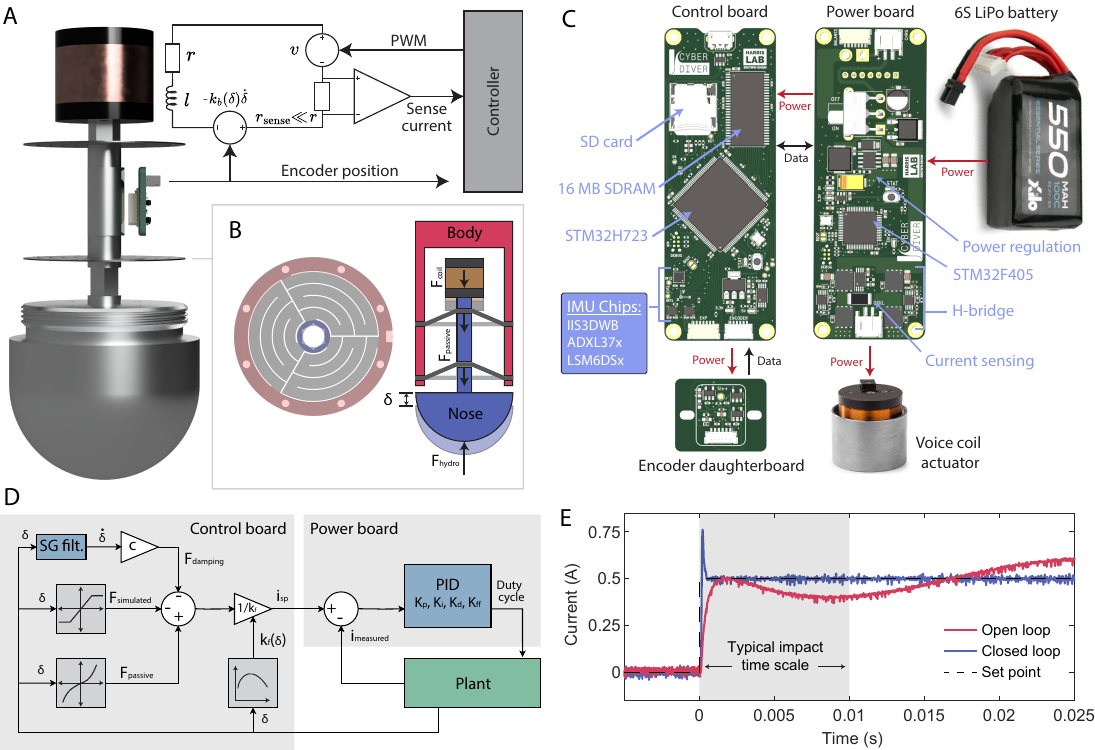}
        \caption{\setlength{\baselineskip}{12pt} \textbf{A voice coil actuator with custom disk flexure bearings is used to simulate arbitrary programmed force-versus-displacement curves using high-speed control electronics and a feedback control mechanism.} (\textbf{A}) The actuator subassembly is shown including the disk flexure bearings, nose hemisphere, linear encoder, and voice coil. The electrical schematic of the actuator is illustrated as well. (\textbf{B}) The motion of the nose relative to the impactor body is constrained by a set of disk flexure bearings that were laser-cut from spring tempered steel. The net force on the impactor nose is the sum of the hydrodynamic force, the passive restoring force of the disk flexures, and the applied force from the voice coil. (\textbf{C}) The CyberDiver electronics consist of three custom printed circuit boards with ARM Cortex-M microcontrollers along with a lithium polymer battery for power during untethered operation and the voice coil actuator. (\textbf{D}) When simulating a programmed structural coupling law, the CyberDiver executes a feedback control loop distributed between the power board and control board. (\textbf{E}) The feedback controller is capable of regulating the coil current even in the presence of back EMF due to relative velocity between the coil winding and magnetic field assembly. The PID controller with aggressively tuned gains can respond to step changes in the set point current on time scales significantly faster than the hydrodynamic loading during impact, avoiding the longer settling time of the open loop response.}
	\label{fig:fig2}
\end{figure}

\begin{figure} 
	\centering
	\includegraphics[width=\textwidth]{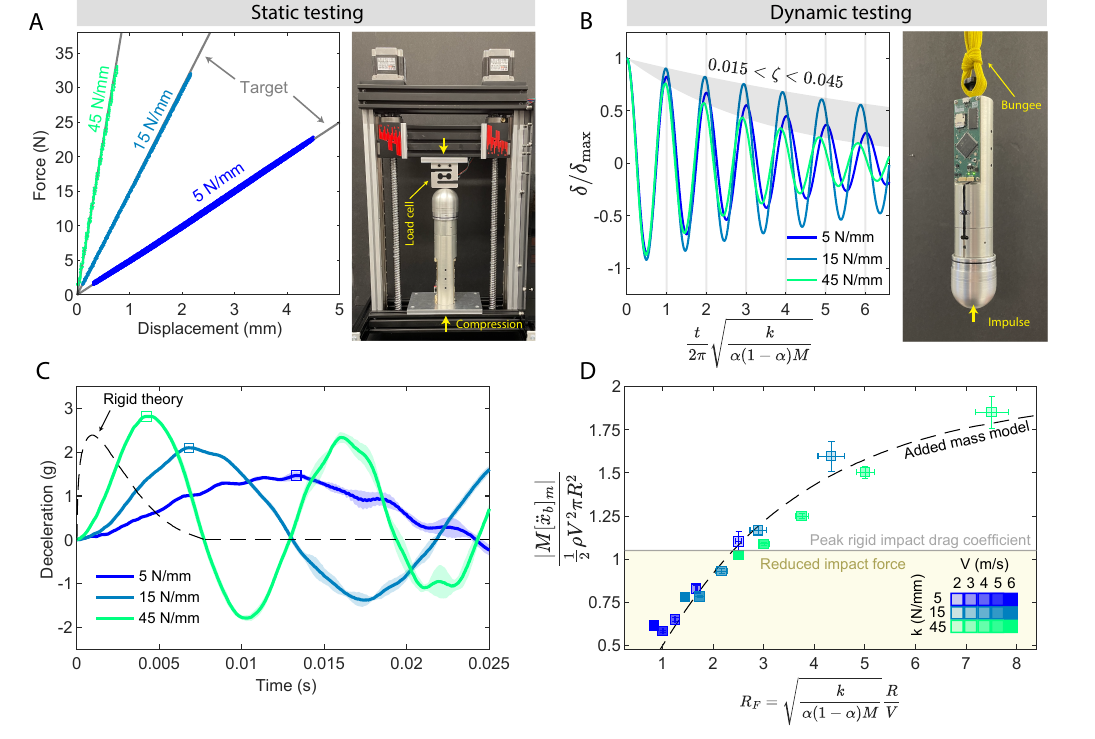}
        \caption{\setlength{\baselineskip}{12pt} \textbf{CyberDiver successfully behaves as linear spring with programmable stiffness in both static and dynamic physical testing, as well as during water-entry experiments.} (\textbf{A}) The force-versus-displacement behavior of the CyberDiver is measured in a compression test machine and matches the target programmed curve. (\textbf{B}) The system is suspended from a bungee and struck with an impulse hammer while the active control is running. The natural frequency of the resulting measured displacement agrees with the programmed linear stiffness. The active feedback system latency and the flexure material damping introduce very little damping since the measured damping ratio of the response varies between 0.015 and 0.045. (\textbf{C}) The acceleration trace during water-entry experiments at 4 m/s reveals the effect of changing the impactor stiffness on the deceleration of the trailing body during impact. By programming a sufficiently small stiffness, the peak deceleration can be reduced substantially compared to the rigid case, reducing the risk of injury to passengers or damage to equipment in engineering applications. The rigid theory comes from the added mass model of Shiffman and Spencer \cite{shiffman1945sphere2} which has been validated thoroughly and used frequently throughout the water-entry literature. The shaded region around the experimental data indicates the standard deviation among 5 experimental replicates. The square markers indicate the peak acceleration in the given experimental conditions. (\textbf{D}) Over a range of velocities and stiffnesses, the normalized peak impact loading measurements collapse with the hydroelastic number, recovering the result from our previous study with an impactor with passive springs \cite{antolik2023shm}. The peak impact loading can be faithfully predicted with an added mass model that takes into account the structural parameters and the geometry of the impactor nose. Replication of previous results from a passive impactor proves that the new cyber-physical system can respond quickly enough during impact and accurately produce the desired structural dynamics.}
	\label{fig:fig3}
\end{figure}

\begin{figure} 
	\centering
	\includegraphics[width=\textwidth]{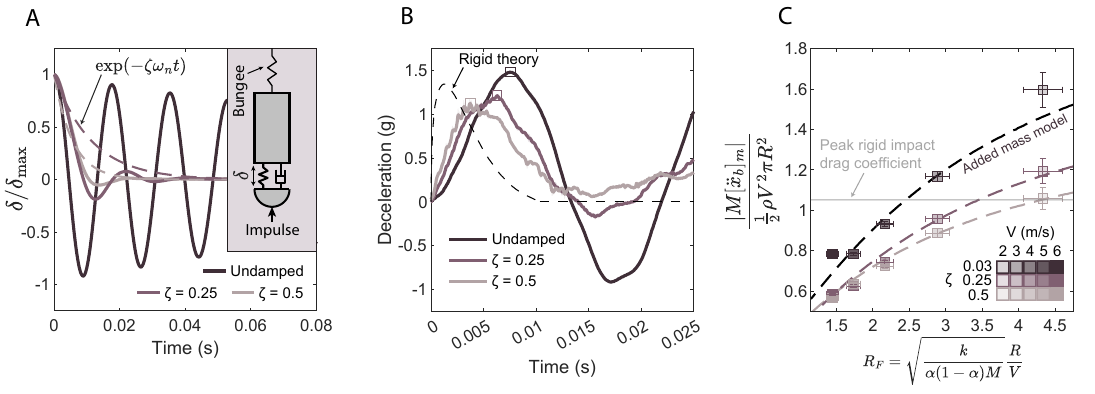}
        \caption{\setlength{\baselineskip}{12pt} \textbf{Introducing a linear damping element by using the active control system further reduces the peak impact loading compared to the undamped case.} (\textbf{A}) Schematic of the programmed structural connection between the impactor nose and trailing body and the setup used to measure the impulse response. The normalized displacements in the ring-down tests are plotted in the solid lines. The dashed lines indicate the expected oscillation envelope decay rate based on the target natural frequency and damping ratio. (\textbf{B}) The programmed damping ratio is evident in the deceleration during impact at speed 3 m/s and linear stiffness 15 N/mm. For these operating parameters, the undamped flexible case experiences higher peak deceleration than the rigid case but the introduction of damping reduces the peak below the rigid case. (\textbf{C}) The introduction of structural damping is beneficial for reducing the dimensionless peak impact loading in the range of hydroelastic numbers tested. In particular, the critical hydroelastic number increases with damping, suggesting that a damped flexible impactor could be used over a wider impact speed range than an undamped impactor without incurring the penalty of a higher impact loading. By introducing a damping term, our added mass model can be used to accurately predict the loading during impact.}
	\label{fig:fig4}
\end{figure}

\begin{figure} 
	\centering
	\includegraphics[width=\textwidth]{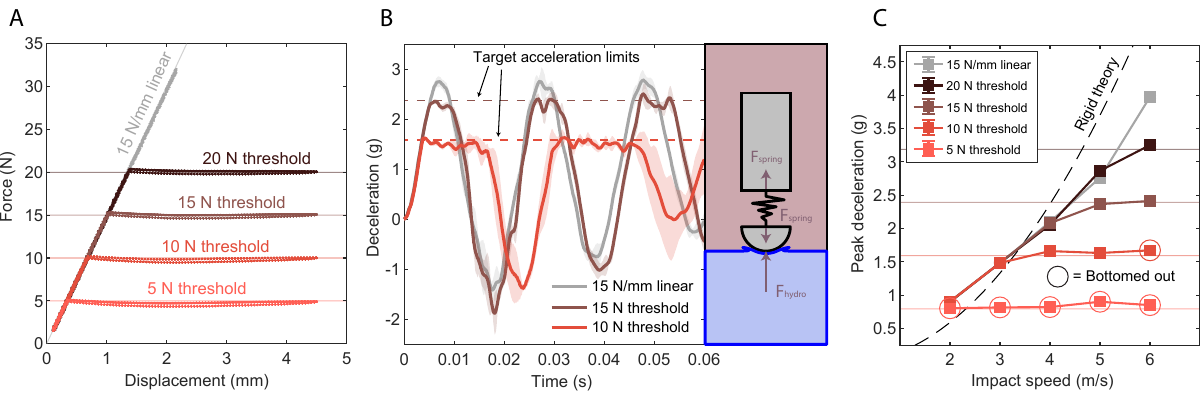}
        \caption{\setlength{\baselineskip}{12pt} \textbf{The active system can be configured with nonlinear structural couplings in order to limit the peak impact loading to a programmable value.} (\textbf{A}) The impactor coupling force is programmed with a linear region near the origin and a threshold beyond which it saturates. The behavior is verified by using a compression testing machine to measure the force-versus-displacement response as the force threshold is changed in software. (\textbf{B}) During the impact experiments shown at 5 m/s, the nonlinear structural coupling removes the peaks in deceleration that would have otherwise surpassed the programmed force threshold. The trailing body only experiences loads from the structural coupling, as shown in the schematic of the programmed structural connection between the impactor nose and trailing body. (\textbf{C}) The peak deceleration is shown as a function of impact speed for various impact experiments and compared to the theoretical rigid result for a sphere. Although the linear spring improves the situation relative to the rigid case at high speeds, the peak dimensional deceleration still continues to increase notably. The force threshold method solves the problem by successfully limiting the peak impact deceleration across all impact speeds tested. The available travel in the impactor motion system is the primary limitation -- cases with too low force thresholds will bottom out. The encircled data points indicate experiments in which the impactor bottoms out before hitting the bottom of the tank (the depth of the water is 70 cm). The reported acceleration in these cases is the maximum measured acceleration while $\delta < \delta_\textrm{max} \approx \textrm{6 mm}$. The reported values are averaged between 5 experimental replicates and the standard deviations are smaller than the marker size.}
	\label{fig:fig5}
\end{figure}

\begin{figure} 
	\centering
	\includegraphics[width=\textwidth]{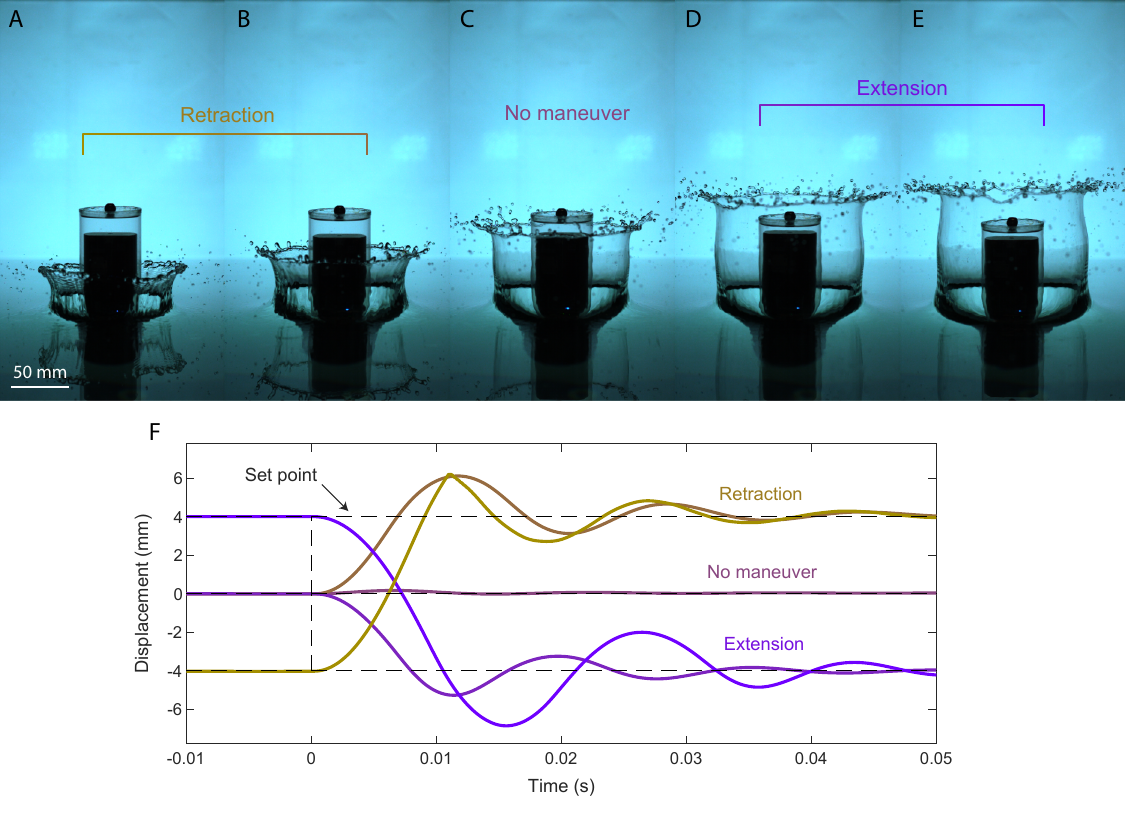}
        \caption{\setlength{\baselineskip}{12pt} \textbf{Active maneuvers coordinated with the moment of impact manipulate the size of the splash crown.} The splash photographs are taken 75 ms after the moment of impact and approximately capture the maximum height of each of the splashes. For these impacts the speed is 2 m/s and a super-hydrophobic coating (Rust-Oleum NeverWet) is applied to the aluminum nose to accentuate the splash formation. (\textbf{A}) The CyberDiver extends the nose to -4 mm displacement during free fall and retracts to 4 mm displacement at impact, minimizing the size of the splash. (\textbf{B}) The nose remains in the neutral position during free fall and retracts to 4 mm displacement at impact. (\textbf{C}) No maneuver is performed and the CyberDiver tries to maintain zero displacement during impact. (\textbf{D}) The nose remains in the neutral position during free fall and extends to -4 mm displacement at impact. (\textbf{E}) The CyberDiver retracts the nose to 4 mm displacement during free fall and extends to -4 mm displacement at impact, maximizing the size of the splash. (\textbf{F}) The actual nose displacement and the set point value in each of the impact cases is shown, with time zero corresponding to the moment of impact with the water's surface. The CyberDiver runs in position control mode and the set point is changed to $\pm$4 mm in the active maneuver cases at time zero. In cases B, C and D, the nose displacement is zero during free fall. In cases A and E, the nose displacement during free fall is $\mp$4 mm, resulting in a maximal change at impact.}
	\label{fig:fig6}
\end{figure}


\clearpage
\bibliography{cyberdiver}

\begin{thebibliography}{10}
\providecommand{\url}[1]{\texttt{#1}}
\expandafter\ifx\csname urlstyle\endcsname\relax
  \providecommand{\doi}[1]{doi:\discretionary{}{}{}#1}\else
  \providecommand{\doi}{doi:\discretionary{}{}{}\begingroup \urlstyle{rm}\Url}\fi

\bibitem{li2025review}
S.~Li, K.~An, W.~Huang, S.~Li, S.~Liu, Research progress of high-speed water entry for trans-media vehicles: State-of-the-art review. \emph{International Communications in Heat and Mass Transfer} \textbf{161}, 108453 (2025).

\bibitem{mcgehee1959nasa}
J.~R. McGehee, M.~E. Hathaway, V.~L. Vaughan~Jr, \emph{Water-landing characteristics of a reentry capsule}, Tech. rep., NASA (1959).

\bibitem{seddon2006aerospace}
C.~Seddon, M.~Moatamedi, Review of water entry with applications to aerospace structures. \emph{International Journal of Impact Engineering} \textbf{32}~(7), 1045--1067 (2006).

\bibitem{pandey2022slamming}
A.~Pandey, J.~Yuk, B.~Chang, F.~E. Fish, S.~Jung, Slamming dynamics of diving and its implications for diving-related injuries. \emph{Science Advances} \textbf{8}~(30), eabo5888 (2022).

\bibitem{chen2017microrobot}
Y.~Chen, \emph{et~al.}, A biologically inspired, flapping-wing, hybrid aerial-aquatic microrobot. \emph{Science Robotics} \textbf{2}~(11), eaao5619 (2017).

\bibitem{li2022airwaterdrone}
L.~Li, \emph{et~al.}, Aerial-aquatic robots capable of crossing the air-water boundary and hitchhiking on surfaces. \emph{Science Robotics} \textbf{7}~(66), eabm6695 (2022).

\bibitem{wang2021auvnumerical}
X.~Wang, Y.~Shi, G.~Pan, X.~Chen, H.~Zhao, Numerical research on the high-speed water entry trajectories of AUVs with asymmetric nose shapes. \emph{Ocean Engineering} \textbf{234}, 109274 (2021).

\bibitem{yan2018auvexperimental}
G.-X. Yan, G.~Pan, Y.~Shi, L.-M. Chao, D.~Zhang, Experimental and numerical investigation of water impact on air-launched AUVs. \emph{Ocean Engineering} \textbf{167}, 156--168 (2018).

\bibitem{von1929impact}
T.~Von~Karman, \emph{The impact on seaplane floats during landing}, Tech. Rep. 321, National Advisory Committee on Aeronautics (1929).

\bibitem{wagner1932stoss}
H.~Wagner, {\"U}ber sto{\ss}-und gleitvorg{\"a}nge an der oberfl{\"a}che von fl{\"u}ssigkeiten. \emph{Zeitschrift f{\"u}r Angewandte Mathematik und Mechanik} \textbf{12}~(4), 192--215 (1932).

\bibitem{abrate2013slamming}
S.~Abrate, {Hull Slamming}. \emph{Applied Mechanics Reviews} \textbf{64}~(6), 060803 (2013).

\bibitem{truscott2014water}
T.~T. Truscott, B.~P. Epps, J.~Belden, Water entry of projectiles. \emph{Annual Review of Fluid Mechanics} \textbf{46}, 355--378 (2014).

\bibitem{jung2021swimming}
S.~Jung, Swimming, flying, and diving behaviors from a unified 2D potential model. \emph{Scientific Reports} \textbf{11}~(1), 15984 (2021).

\bibitem{baldwin1971vertical}
J.~L. Baldwin, \emph{Vertical water entry of cones}, Tech. Rep. NOLTR-71-25, Naval Ordnance Lab, White Oak, MD (1971).

\bibitem{chang2016seabirds}
B.~Chang, \emph{et~al.}, How seabirds plunge-dive without injuries. \emph{Proceedings of the National Academy of Sciences} \textbf{113}~(43), 12006--12011 (2016).

\bibitem{gilbert2023wedge}
C.~Gilbert, J.~Gilbert, M.~J. Javaherian, Water entry of a flexible wedge: How flexural rigidity influences spray root and pressure wave propagation. \emph{Phys. Rev. Fluids} \textbf{8}, 090502 (2023).

\bibitem{belden2024sphereflat}
J.~Belden, N.~Speirs, A.~Hellum, G.~Loubimov, T.~T. Truscott, Water impact: When a sphere becomes flat. \emph{Phys. Rev. Lett.} \textbf{133}, 034002 (2024).

\bibitem{wu2020water}
Z.~Wu, \emph{et~al.}, Water entry of slender segmented projectile connected by spring. \emph{Ocean Engineering} \textbf{217}, 108016 (2020).

\bibitem{boom2023water}
B.~Boom, T.~Truscott, F.~Fish, A.~Summers, E.~Habtour, Water entry dynamics of avian inspired divers, in \emph{Smart Materials, Adaptive Structures and Intelligent Systems} (American Society of Mechanical Engineers), vol. 87523 (2023), p. V001T06A001.

\bibitem{antolik2023shm}
J.~T. Antolik, J.~L. Belden, N.~B. Speirs, D.~M. Harris, Slamming forces during water entry of a simple harmonic oscillator. \emph{Journal of Fluid Mechanics} \textbf{974}, A23 (2023).

\bibitem{miller1996cps}
S.~N. Miller, \emph{Measurement of vortex-induced oscillations of marine cables using feedback with explicit structural modeling}, Master's thesis, Massachusetts Institute of Technology (1996).

\bibitem{hover1997vivmarinecables}
F.~Hover, S.~Miller, M.~Triantafyllou, Vortex-induced vibration of marine cables: experiments using force feedback. \emph{Journal of Fluids and Structures} \textbf{11}~(3), 307--326 (1997).

\bibitem{hover1998oscillatingcylinders}
F.~S. Hover, A.~H. Techet, M.~S. Triantafyllou, Forces on oscillating uniform and tapered cylinders in cross flow. \emph{Journal of Fluid Mechanics} \textbf{363}, 97–114 (1998).

\bibitem{mackowski2011cps}
A.~W. Mackowski, C.~H. Williamson, Developing a cyber-physical fluid dynamics facility for fluid–structure interaction studies. \emph{Journal of Fluids and Structures} \textbf{27}~(5), 748--757 (2011), iUTAM Symposium on Bluff Body Wakes and Vortex-Induced Vibrations (BBVIV-6).

\bibitem{lee2011viv}
J.~Lee, N.~Xiros, M.~Bernitsas, Virtual damper–spring system for VIV experiments and hydrokinetic energy conversion. \emph{Ocean Engineering} \textbf{38}~(5), 732--747 (2011).

\bibitem{onoue2016cpspitchingplate}
K.~Onoue, K.~S. Breuer, Vortex formation and shedding from a cyber-physical pitching plate. \emph{Journal of Fluid Mechanics} \textbf{793}, 229–247 (2016).

\bibitem{zhu2020cpsflume}
Y.~Zhu, Y.~Su, K.~Breuer, Nonlinear flow-induced instability of an elastically mounted pitching wing. \emph{Journal of Fluid Mechanics} \textbf{899}, A35 (2020).

\bibitem{fagley2016aeroelastic}
C.~Fagley, J.~Seidel, T.~McLaughlin, Cyber-physical flexible wing for aeroelastic investigations of stall and classical flutter. \emph{Journal of Fluids and Structures} \textbf{67}, 34--47 (2016).

\bibitem{fan2019towtank}
D.~Fan, \emph{et~al.}, A robotic Intelligent Towing Tank for learning complex fluid-structure dynamics. \emph{Science Robotics} \textbf{4}~(36), eaay5063 (2019).

\bibitem{zhong2021fish}
Q.~Zhong, \emph{et~al.}, Tunable stiffness enables fast and efficient swimming in fish-like robots. \emph{Science Robotics} \textbf{6}~(57), eabe4088 (2021).

\bibitem{lee2022mnn}
R.~H. Lee, E.~A.~B. Mulder, J.~B. Hopkins, Mechanical neural networks: Architected materials that learn behaviors. \emph{Science Robotics} \textbf{7}~(71), eabq7278 (2022).

\bibitem{lee1981plummeting}
D.~N. Lee, P.~E. Reddish, Plummeting gannets: A paradigm of ecological optics. \emph{Nature} \textbf{293}~(5830), 293--294 (1981).

\bibitem{brierley2001gannets}
A.~S. Brierley, P.~G. Fernandes, Diving depths of northern gannets: Acoustic observations of Sula Bassana from an autonomous underwater vehicle. \emph{The Auk} \textbf{118}~(2), 529--534 (2001).

\bibitem{sharker2019water}
S.~I. Sharker, S.~Holekamp, M.~M. Mansoor, F.~E. Fish, T.~T. Truscott, Water entry impact dynamics of diving birds. \emph{Bioinspiration \& Biomimetics} \textbf{14}~(5), 056013 (2019).

\bibitem{gregorio2023air}
E.~Gregorio, E.~Balaras, M.~C. Leftwich, Air cavity deformation by single jointed diver model entry bodies. \emph{Experiments in Fluids} \textbf{64}~(11), 168 (2023).

\bibitem{rohilla2024manu}
P.~Rohilla, \emph{et~al.}, Mastering the Manu: How humans create large splashes. \emph{bioRxiv} pp. 2024--12 (2024).

\bibitem{desilva2015sensors}
C.~W. De~Silva, \emph{Sensors and actuators: Engineering system instrumentation} (CRC press) (2015).

\bibitem{schafer2011savitzky}
R.~W. Schafer, What is a savitzky-golay filter?[lecture notes]. \emph{IEEE Signal processing magazine} \textbf{28}~(4), 111--117 (2011).

\bibitem{shiffman1945sphere1}
M.~Shiffman, D.~C. Spencer, \emph{The force of impact on a sphere striking a water surface: approximation by the flow about a lens}, Tech. Rep. AMG-NYU-105, New York University, Courant Institute of Mathematical Sciences (1945).

\bibitem{shiffman1945sphere2}
M.~Shiffman, D.~C. Spencer, \emph{The force of impact on a sphere striking a water surface: second approximation}, Tech. Rep. AMG-NYU-133, New York University, Courant Institute of Mathematical Sciences (1945).

\bibitem{shiffman1947lens}
M.~Shiffman, D.~C. Spencer, The flow of an ideal incompressible fluid about a lens. \emph{Quarterly of Applied Mathematics} \textbf{5}~(3), 270--288 (1947).

\bibitem{moghisi1981experiments}
M.~Moghisi, P.~T. Squire, An experimental investigation of the initial force of impact on a sphere striking a liquid surface. \emph{Journal of Fluid Mechanics} \textbf{108}, 133–146 (1981).

\bibitem{yan2022qzs}
G.~Yan, \emph{et~al.}, Nonlinear compensation method for quasi-zero stiffness vibration isolation. \emph{Journal of Sound and Vibration} \textbf{523}, 116743 (2022).

\bibitem{chai2024qzs}
Y.~Chai, J.~Bian, M.~Li, A novel quasi-zero-stiffness isolation platform via tunable positive and negative stiffness compensation mechanism. \emph{Nonlinear Dynamics} \textbf{112}~(1), 101--123 (2024).

\bibitem{meymian2018flexures}
N.~Z. Meymian, \emph{et~al.}, An optimization method for flexural bearing design for high-stroke high-frequency applications. \emph{Cryogenics} \textbf{95}, 82--94 (2018).

\bibitem{achanur2014shapefactor}
M.~Achanur, Shape factor optimization and parametric analysis of spiral arm flexure bearing through finite element analysis studies. \emph{International Journal of Engineering Research} \textbf{3}~(4) (2014).

\bibitem{teo2015flexurepositioning}
T.~J. Teo, G.~Yang, I.~Chen, A flexure-based electromagnetic nanopositioning actuator with predictable and re-configurable open-loop positioning resolution. \emph{Precision Engineering} \textbf{40}, 249–260 (2015).

\bibitem{antolik2025impack}
J.~T. Antolik, D.~M. Harris, harrislab-brown/IMpack: IMpack (v1.0) (2025), \doi{https://doi.org/10.5281/zenodo.15014096}.

\bibitem{ogata2009modern}
K.~Ogata, \emph{Modern control engineering} (Pearson) (2009).

\end{thebibliography}
\bibliographystyle{sciencemag}


\section*{Acknowledgments}

We thank Oskar Weigl at ODrive for his valuable insights into the design of the H-bridge modulation scheme. 

\paragraph*{Funding:}
J.T.A., E.A.S., and D.M.H. acknowledge funding from the Office of Naval Research (ONR N00014-21-1-2816) and the NASA-Rhode Island Space Grant Consortium. J.L.B. acknowledges funding from the Naval Undersea Warfare Center In-House Laboratory Independent Research program, monitored by Dr E. Magliula.

\paragraph*{Author contributions:}

J.L.B., D.M.H., and J.T.A. conceived the study and secured funding. J.T.A., J.L.B., and D.M.H. designed the study and discussed results. J.T.A. and E.A.S. designed and constructed the CyberDiver. J.T.A. performed the experiments. J.T.A. and D.M.H. analyzed data, developed the mathematical model, and wrote the draft of the manuscript. All authors provided feedback on the manuscript and approved the final version. D.M.H. supervised the overall project.

\paragraph*{Competing interests:}

There are no competing interests to declare.

\paragraph*{Data and materials availability:}

Raw experimental data and movies associated with this article, scripts used in processing the data and running the theoretical model, mechanical and electrical design files, and firmware source code for the CyberDiver are available at:\\ \url{https://doi.org/10.5281/zenodo.15079391}.


\subsection*{Supplementary materials}
Materials and Methods\\
Supplementary Text\\
Figs. S1 to S5\\
Table S1\\
References \textit{(45-\arabic{enumiv})}\\ 
Movies S1 to S4\\


\newpage


\renewcommand{\thefigure}{S\arabic{figure}}
\renewcommand{\thetable}{S\arabic{table}}
\renewcommand{\theequation}{S\arabic{equation}}
\renewcommand{\thepage}{S\arabic{page}}
\setcounter{figure}{0}
\setcounter{table}{0}
\setcounter{equation}{0}
\setcounter{page}{1}


\begin{center}
\section*{Supplementary Materials for\\ \scititle}

John~T.~Antolik,
Eli~A.~Silver,
Jesse~L.~Belden,
Daniel~M.~Harris$^\ast$\\ 
\small$^\ast$Corresponding author. Email: daniel\_harris3@brown.edu\\
\end{center}

\subsubsection*{This PDF file includes:}
Materials and Methods\\
Supplementary Text\\
Figures S1 to S5\\
Table S1\\
Captions for Movies S1 to S4\\

\subsubsection*{Other Supplementary Materials for this manuscript:}
Movies S1 to S4\\

\newpage


\subsection*{Materials and Methods}

\subsubsection*{Disk flexure design} \label{sec:flexures_design}

A central element of the motion subsystem of the CyberDiver is the disk flexure bearing set that constrains motion between the impactor body and the hemispherical nose. Flexure elements are used rather than traditional sliding bearings because the measurements of impact loading are sensitive and can be corrupted by the effects of friction. Furthermore sliding friction is difficult to characterize and would need to be dynamically accounted for in the cyber-physical system which would be computationally intensive for the embedded hardware. By using only flexural coupling elements (and a contactless measurement technique to determine the nose displacement) the passive coupling force can be modeled as a function of instantaneous displacement alone with very high accuracy. A similar construction was used by \cite{lee2022mnn} to successfully implement an array of cyber-physical springs. In our system we use a pair of thin circular disk flexures to couple the impactor nose and body. The outer ring of the flexure is rigidly clamped to the impactor body and the inner ring is rigidly clamped to a central shaft that connects to the hemispherical nose, as shown in Figure \ref{fig:fig2}(B). The hexagonal cutout in the center is used to index the encoder glass scale mounting hardware and align it with the read head. The flexures deform out of plane as the impactor nose is displaced. By using a pair of flexures with significant axial separation, the overall structure obtains high bending stiffness relative to its axial stiffness. The axial and radial stiffness of the flexure can be manipulated through the design of the cut profile between the inner and outer rings. Disk flexures in the literature use Archimedes spirals which can be modified by a shape factor \cite{meymian2018flexures, achanur2014shapefactor} or folded beam based cutouts \cite{teo2015flexurepositioning}. Our disk flexures are based on the folded beam design and are laser-cut from AISI 1075 tempered spring steel sheet with 0.38 mm thickness. We use the Autodesk Fusion 360 application programming interface (API) with custom Python scripts to parametrically generate disk flexures of the aforementioned types for physical testing. The design was also informed by nonlinear static simulations and linear modal analysis simulations conducted in Autodesk Fusion 360. The flexures must remain in the elastic material regime throughout the travel of the impactor nose. Since the intent is to construct an impactor with a single axial degree of freedom, we wish to maximize the ratio of the radial stiffness to the axial stiffness. Additionally, the force response in the axial direction should be as close to linear as possible so that the cyber-physical system is well behaved. For our system, a folded beam disk flexure with three-fold rotational symmetry and three folds per beam satisfies these constraints. The linear term in a cubic fit ($F_p(\delta) =k_1 \delta + k_3 \delta^3$) to the force-versus-displacement measurement of our disk flexure pair is $k_1=4.01$ N/mm and the magnitude of the cubic term is $k_3=0.0091$ N/mm$^3$.  

\subsubsection*{CyberDiver mechanical design}

The overall form of the CyberDiver is a slender axisymmetric body with a small nose piece that contacts the water initially, preceding a larger trailing body which is attached via the structural coupling that can deform in the axial direction. In order to waterproof the electronics, the moving joint between the nose and body needs to be sealed. If the joint itself were sealed with a rubber membrane or sliding O-ring, there would be additional frictional contributions to the coupling force that would need to be carefully characterized and compensated for in the cyber-physical control loop. An alternative solution that we implement for the CyberDiver is to affix a thin transparent plastic shell to the nose which surrounds and seals the entire system. Thus there is no additional contribution to the coupling force between the nose and body because there is sufficient clearance between the body and shell for the structure to vibrate freely during impact. The sealing shell defines the overall envelope in which the sensor, actuator, bearings and electronics must fit. The primary structure of the CyberDiver body is the main housing machined from ASTM 6061-T6 aluminum. This material was selected for its high strength and high elastic wave speed with the intent of well-separating higher modes of vibration within the body from the fundamental axial mode of interest. The main housing features a pocket for the battery which is clamped in place with rubber-tipped set screws and sandwiched between the long rectangular control and power printed circuit boards (PCBs) on either side. The battery pocket ends are extended to accommodate the battery cables and the power and data wires which connect the two PCBs. The end of the main housing is a precision bore which contains the magnetic housing of the voice coil actuator. The magnetic housing is fastened with a single \#8 screw from inside the battery pocket. Eight \#2-56 holes are tapped around the perimeter of the field assembly bore to receive the long screws that clamp the outer perimeter of the disk flexure bearing stack. The intermediate housing that maintains the axial separation between the disk flexures is split into two halves which aids assembly and allows access for adjustment of the encoder alignment. The encoder PCB is mounted to one of the halves of the intermediate housing and must be precisely aligned to the glass scale which moves axially with the nose. There are two opposing set screws in the intermediate housing for adjusting the lateral position of the encoder PCB. The spacing between the PCB and scale can be adjusted by adding shims behind the PCB; custom 0.1 mm thickness shims were fabricated using the solder stencil services of a PCB prototyping supplier. There is a clearance hole in the housing that mounts the encoder PCB so that a probe can be attached to a test point on the back of the encoder board with an analog alignment signal from the encoder chip. The tolerance for the spacing between the scale and encoder is $\pm$0.25 mm and the tolerance for the lateral alignment is $\pm$0.2 mm. The required angular alignment in all axes is 1.5 degrees. Alignment in the direction of motion is not necessary because the axial displacement offset corresponding to the neutral position of the flexure bearings is measured and compensated in software. When the encoder is properly aligned, the analog tuning signal will produce a sine wave as the encoder is displaced relative to the scale whose amplitude is between 1.43 V and 2.38 V peak-to-peak. The encoder scale holder provides the axial separation between the inner rings of the disk flexures and indexes into the hexagonal cutouts in the flexures in order to maintain angular alignment between the encoder and scale. The central shaft threads into the coil of the actuator, passes through the flexure bearings and scale holder, and receives a coupling nut on the other end that clamps the inner rings of the flexure stack together. The central shaft must be able to rotate relative to the scale holder so that the coil can be aligned independently from the glass scale. When installed, the coil is rotated so that its wires exit directly through the allotted wire slot in the main housing -- otherwise they would be free to rub against the housing and produce unwanted friction. After connecting the coil to the power board, the wires are laid into the slot in the external surface of the main housing and are clamped in place with a laser-cut piece of plastic and two screws. There is another slot on the opposite side of the main housing to hold the wires which connect the encoder to the control board. The hemispherical nose is composed of three pieces with a hollow interior to reduce mass. There are threaded connections with O-ring seals between the hollow hemisphere, a flat intermediate plate, and the ring that provides an easily removable connection to the sealing shell. The flat intermediate nose plate has mounting provisions for an accelerometer daughter board so that the nose acceleration could be directly measured in the future. The nose assembly is designed to have low mass since it sets the minimum $\alpha$ parameter that we can achieve in experiments -- higher $\alpha$ can be achieved in the future by adding mass in the hollow nose region, but $\alpha$ was not varied in the present study. The shell is a transparent PETG tube with an end cap. Three locking tabs are cut around the open end of the shell so that it can be attached to and easily removed from the nose ring. A steel ball with a threaded hole is fastened to the end of the shell so that the CyberDiver can be dropped from an electromagnet.

\subsubsection*{Control electronics}

In order for the control electronics to fit within the envelope prescribed by the mechanical system, we designed custom PCBs that gather and log data from the sensors, drive the voice coil actuator, and perform the active control calculations. The responsibilities of the electronics system are split between two circuit boards which we refer to as the control board and the power board. The power board features the power electronics necessary to drive the voice coil actuator. The control board handles data logging, reading data from the accelerometers and encoder, and running the overall application code. The two boards communicate via the Serial Peripheral Interface (SPI). By separating tasks between two boards which communicate over a standardized interface, the system is effectively decoupled and the design of one of the boards may be changed without influencing the design of the other board. This strategy minimizes risk in the design process which requires a lot of customized circuitry. It also grants more overall computational power for running the control since the calculations are split between two different microcontrollers. The downside is that the communication link between the boards limits the controller bandwidth since data must be transferred between them at each update of the controller. Throughout the design and prototyping phases of the project, individual sub-circuits were tested and validated before the final boards for the CyberDiver were designed. A prototype of the power electronics for the voice coil was implemented using the standard Arduino shield format so that it could be tested with a number of different microcontrollers on development boards. The encoder chip implementation was tested by using the manufacturer's development board and a custom standalone signal converter board. Crucially, given that acceleration is the primary measurement of our experiments, the inertial measurement unit (IMU) chip implementation and data logging system was tested on a compact standalone board that could be mounted in a rigid impactor alongside a commercial IMU and directly evaluated. This board was highly effective and useful in ongoing impact experiments so it evolved into its own open source project, the IMpack \cite{antolik2025impack}.

The control board implements the STM32H723 microcontroller unit (MCU) with 550 MHz clock speed and double precision floating point unit due to its high computational power and wide array of peripherals including quadrature encoder timer hardware, integrated full-speed Universal Serial Bus (USB) layer, SPI hardware, and support for Secure Digital (SD) cards and external random access memory (RAM). The implementation of the MCU is straightforward, requiring an external crystal oscillator as a clock source, power supply filtering, provisions for boot mode selection and a programming connector. We use the Serial Wire Debug (SWD) protocol via the STLINK-V3MINIE probe to program and debug firmware on the CyberDiver. The control board receives 5 V power from the power board which it regulates to 3.3 V for the microcontroller and associated peripherals. The control board features a micro USB connector so that it can be controlled from a host interface on a desktop computer. The CyberDiver enumerates as a USB virtual communication port and can receive a set of serial commands or report data while calibrating and tuning the device. The control board features a power multiplexer so that it can be powered solely over the USB port during firmware development or testing without the fully assembled impactor. A compact micro SD card slot (Hirose DM3CS-SF) is used to log experimental data via the Secure Digital/Multi-Media Card (SDMMC) peripheral of the microcontroller. SD cards have unpredictable write times due to factors such as wear leveling and the maximum write latency for the micro SD card used (SanDisk Industrial 8 GB) is 250 ms. The CyberDiver logs several parameters at frequencies up to 50 kHz resulting in an overall data rate of 2.2 MB/s. Given the SD card latency and the double buffering scheme used to acquire data, the previous 1.1 MB of data must be always stored in memory in order to avoid data loss. Since this exceeds the internal RAM of the MCU, we extend the available memory using the microcontroller's flexible memory controller (FMC) peripheral and a synchronous dynamic random access memory (SDRAM) chip (W9812G6KH-6) with 16 MB capacity. This allows up to an entire impact's worth of data to be buffered externally on the SDRAM chip without complications from the SD card write latency. The length variation of the routed memory bus lines on the PCB is within $\pm$23 mm, assuring that propagation delay variance of the memory signals is within 2\% of the clock period of the 100 MHz memory clock. An array of MEMS accelerometers which communicate data via SPI to the MCU is placed near one of the board mounting holes for greater mechanical rigidity. The IMU chips used include the IIS3DWB, ADXL373, and the LSM6DS3 with the primary difference being a trade-off between sampling rate and measurement range. We use the IIS3DWB to gather data during our impact experiments since it has the highest sampling rate of 26.6 kHz. It contains a 16-bit analog-to-digital converter, maximum measurement range of $\pm$16 g, and low acceleration noise density in the impact direction of 75 $\mu \textrm{g} / \sqrt{\textrm{Hz}}$. Finally the control board features connectors for the external sensors including the encoder daughter board and an expansion port for a future accelerometer on the impactor nose.

The power board implements the comparatively simpler STM32F405 microcontroller with 168 MHz core clock speed for standalone closed loop control of the voice coil. It features connectors for the battery power cable and the charge connector that is used to evenly charge the cells of the lithium polymer battery pack. When the main power switch is in the off position, the power and charge connectors pass through directly to small connectors on the top of the board so the CyberDiver can be charged using an off-the-shelf external charger without any disassembly. The nominal battery voltage of 25.5 V is converted to 5 V using a buck converter circuit based on the TPS5430 chip to power the logic sections while the full battery voltage is applied to the H-bridge. An additional 12 V supply is generated using a linear regulator to power the gate driver components in the H-bridge. The switching elements of the H-bridge are four N-channel metal–oxide–semiconductor field-effect transistors (MOSFETs, part number SIR182DP) which can support drain-source voltage up to 60 V and maximum continuous drain current of 117 A. We use LM5106 half-bridge gate drivers to drive each set of high and low side MOSFETs. The gate drivers feature a bootstrap circuit to generate a sufficiently high voltage to switch the gate of the N-channel transistors based on a logic level signal input from the microcontroller. Additionally, the gate drivers automatically insert an adjustable dead time interval when switching the MOSFETs to prevent short circuits through the bridge due to the finite switching time of the MOSFETs. The dead time is configured by applying a known resistance to the dead time pin of the gate driver -- our circuit uses 150 ns of dead time. We also include 1N5819 flyback diodes in the H-bridge to eliminate the voltage spike from the coil inductor when the supplied current is interrupted by switching the power transistors. In order to implement closed loop control of the coil current, we use an inline 6 m$\Omega$ shunt resistor along with the INA240 current sense instrumentation amplifier. Inline current sensing (in which the sense element is directly in series with the coil) is used here as opposed to other strategies such as high-side or low-side sensing because the CyberDiver needs to know the coil current at all times in order to quickly counteract the effects of back EMF and maintain the desired coil force. With high-side or low-side sensing schemes, the coil current can only be sampled at particular times within the H-bridge modulation cycle. The downside of inline current sensing is that the sense amplifier is exposed to large amplitude fluctuations in the common mode, or average, voltage at the sense leads. When the H-bridge changes the direction of the applied voltage, the common mode voltage at the sense amplifier quickly changes from near zero volts to the battery voltage, or vice versa. For many lower-performance sense amplifiers, this common mode transient carries through to the amplifier's output even though only the differential voltage measurement is desired. The INA240 sense amplifier is chosen because it features a very high common mode rejection ratio (CMRR) of 93 dB at 50 kHz which preserves a stable differential output reading even in the presence of large common mode transients. Nevertheless, the common mode transients are not completely attenuated so the control scheme we implement attempts to sample the current only when the common mode voltage is most stable. This scheme is discussed in detail in a later section. The shunt resistance, sense amplifier gain and amplifier configuration are selected in such a way that the full scale coil current readings from -5.5 A to +5.5 A map linearly to a sense amplifier output ranging from 0 to 3.3 V. The sense amplifier output voltage feeds into the 12 bit analog-to-digital converter (ADC) integrated in the MCU after passing through a simple first order low pass filter with 32 kHz corner frequency. The MCU switches each half of the H-bridge using one channel of a hardware timer configured to generate a center-aligned pulse width modulation (PWM) signal.  

High-resolution, high-speed, contactless position sensing within the space constraints of the integrated system was a significant challenge addressed by designing a custom encoder daughter board based on the Celera CE300-40 encoder chip. This encoder chip has a footprint of only 7 by 11 mm and measures with one micron accuracy at speeds up to 14 m/s. The chip comes in a land grid array style package meaning that the solder pads are on the back of the component and inaccessible to hand soldering. In order to use this encoder chip with our control electronics, we designed a custom PCB that takes in a 5 V supply and provides the requisite power conditioning circuitry for the encoder and signal conversion so that the encoder outputs can be fed directly into the main microcontroller. All of the peripheral components are on the back of the daughter board and we manually solder the encoder chip to the front of the board using custom solder paste stencils and a hot air soldering station. The encoder chip outputs differential RS-422 quadrature signals along with a pulse for the index on the scale. We use MAX3280 receiver components to convert them to single ended signals and level shift them to 3.3 V which can be directly accepted by the MCU on the control board. The STM32H723 microcontroller features timer peripherals that can accept quadrature encoder signals so the MCU can track the nose displacement reliably without any intervention from the central processing unit (CPU). 

\subsubsection*{Control software}

The firmware for the CyberDiver is written in C using the STM32 hardware abstraction layer (HAL) for interfacing with the hardware peripherals. Separate applications are developed for the power board and the control board with a common interface for the boards to share commands or data over the SPI communication channel. The applications are developed using an object-oriented approach in which the data and functions associated with particular subsets of the application functionality are encapsulated into classes. This approach allows the complicated behavior of the overall application to be composed of relatively simpler pieces whose behavior can be individually specified, developed and tested. Since C is generally not an object-oriented programming language, we only rely on simple object-oriented design concepts such as encapsulation and make no attempt to utilize deeper elements like inheritance or polymorphism. The overall structure of both the control board and power board applications consists of a main loop for tasks with no particular time requirement and a time critical ``thread'' based on a high-priority timer interrupt in which the real-time controllers are updated. 

\begin{figure}
    \centering
    \includegraphics[width=\textwidth]{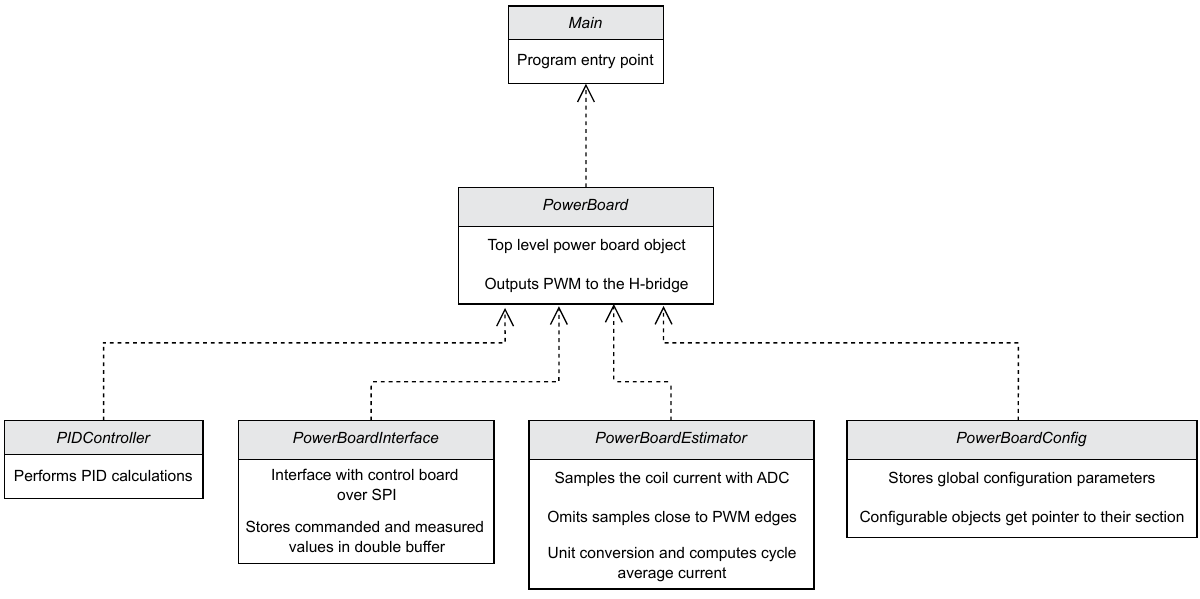}
    \caption{\setlength{\baselineskip}{12pt} \textbf{Class diagram of the firmware running on the CyberDiver power board.} The overall organization of the power board firmware is shown in the diagram. Each block represents a class with its name labeled at the top. The salient data that the class contains and the functions that it performs are outlined within the box. The arrows are an indicator of dependency. In other words, if an arrow points from one class to another the class pointed to has knowledge of the class at which the arrow originates.}
    \label{fig:fig7}
\end{figure}

The power board application is relatively simple and its overall structure is illustrated in Figure \ref{fig:fig7}. It can exist in either the idle state, in which it disables the H-bridge and waits to receive configuration parameters, or the active state, in which it listens for SPI commands for the coil current set point value, continuously samples the coil current, updates its PID controller, and adjusts the duty cycle of the PWM signal that it sends to the H-bridge. Given the considerable computational power of the MCU, the power board PID control calculations are performed in the 32-bit floating point domain and update at a rate of 50 kHz. However, data and commands are transferred between the boards as 16-bit fixed point numbers to improve the communication rate. An overarching design goal of the controller software is to minimize the delay between sampling the sensors and adjusting the actuator output at each control cycle in order to more closely approximate the continuous-time system. Since the overall controller is split between two separate microcontrollers, the serial communication between them significantly contributes to the computational delay in the controller. As such, in addition to compressing the data to be transferred between the boards, we also carefully optimize the code responsible for the SPI serial communication. SPI follows a controller-peripheral architecture in which the controller device manages the communication, including generating the clock signal, and the peripheral device merely responds. It is a full duplex protocol meaning that data can be transferred in both directions simultaneously. In our implementation the power board is the peripheral device and the control board serves as the SPI controller. The SPI clock is configured with a frequency of 13.75 MHz which we determine to be near the upper limit for robust communication before the signal integrity starts to decline. When the voice coil is active, the power board SPI hardware is configured to operate with the direct memory access (DMA) hardware in circular mode. This means that the SPI hardware will continuously listen for communications and the received data will automatically be transferred via DMA to a prescribed address in memory without any intervention from the CPU. As a result, the data necessary for updating the voice coil controller can be accessed with minimal overhead. After receiving each packet of data, we update the destination memory address for the next for the packet of data received over SPI while the PID controller maintains a reference to the original memory address. This scheme avoids the possibility of a so-called data race condition in which the PID controller is partway through accessing the old data when a new packet arrives, thus corrupting the value. After each new packet of data, we swap the memory addresses, thereby implementing a so-called double buffer (or ping-pong buffer). The data itself contains two 16-bit numbers. In each transaction, the new coil current set point and the most recent relative velocity measurement are sent from the control board to the power board. Simultaneously, the most recent measured current and PWM duty cycle are sent from the power board to the control board to be logged. The measured relative velocity is transmitted to the power board so that it can be used in a feed-forward term of the control that proactively mitigates the effects of back EMF. However, we find in practice that this is not necessary to achieve the desired performance during water entry. The speed of the SPI transaction can also be optimized at the control board side. In particular, the control board contributes overhead time to the SPI transaction since it is responsible for managing the communication. In order to minimize the overhead, we utilize the SPI hardware first-in-first-out (FIFO) buffer on the STM32H7 and write our own subroutine that directly manipulates the SPI registers to fill data into the buffer and transmit it. There are other strategies that use interrupts or DMA to avoid blocking the CPU while the data is transmitted, but the trade off for offloading the CPU is greater overhead in setting up each transaction. For a system like ours with frequent but short communications, we find our implementation to be the best option. Measurements with a logic analyzer reveal that the full communication between the boards at each update cycle takes approximately 3.8 $\mu$s. 

\begin{figure}
    \centering
    \includegraphics[width = 0.75\textwidth]{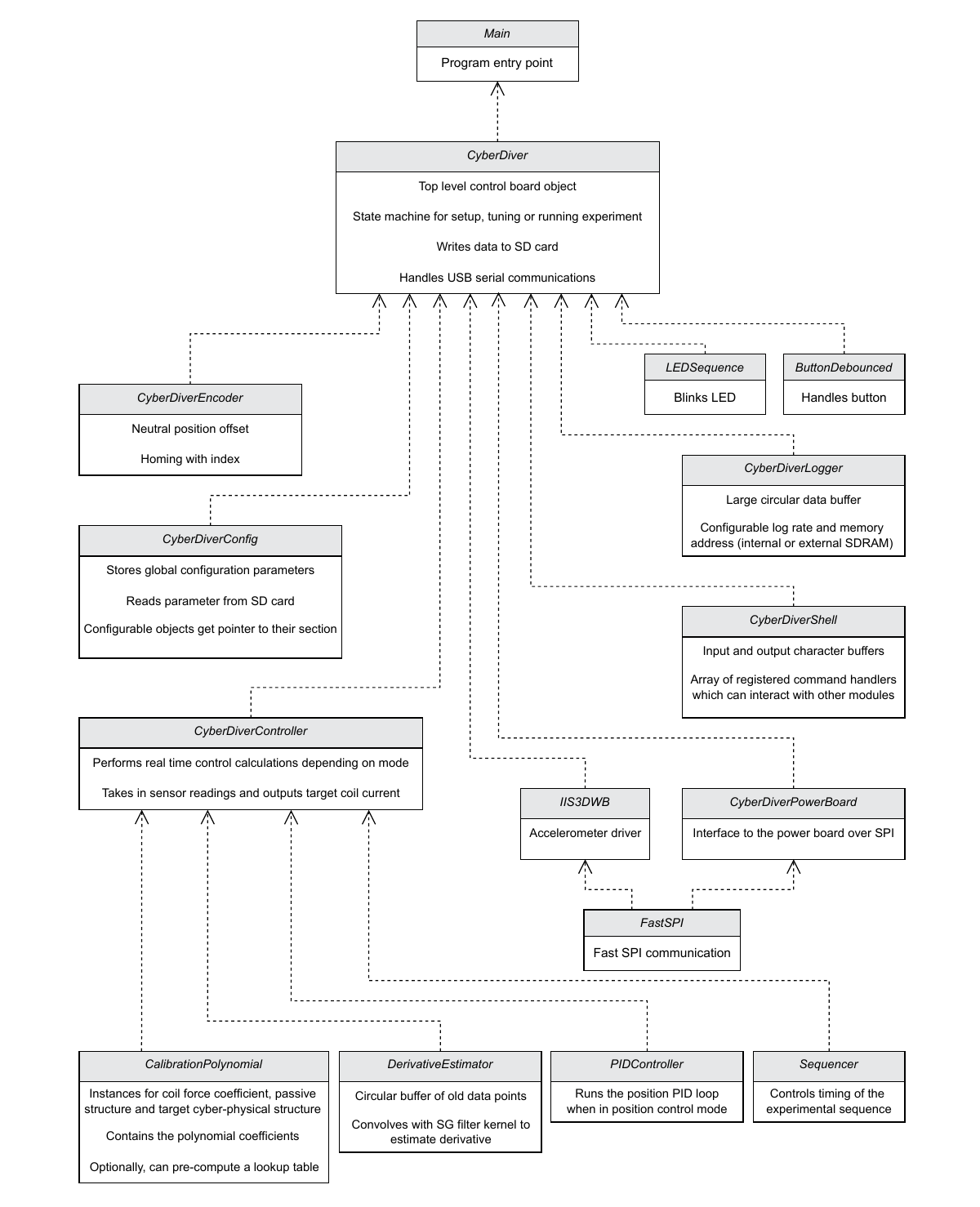}
    \caption{\setlength{\baselineskip}{12pt} \textbf{Class diagram of the firmware running on the CyberDiver control board.} The overall organization of the control board firmware is shown in the diagram. Each block represents a class with its name labeled at the top. The salient data that the class contains and the functions that it performs are outlined within the box. The arrows are an indicator of dependency. In other words, if an arrow points from one class to another the class pointed to has knowledge of the class at which the arrow originates.}
    \label{fig:fig8}
\end{figure}

The control board application whose overall structure is presented in Figure \ref{fig:fig8} relies on a state machine at the highest level which orchestrates its overall operating mode. At startup, the control board enters a configuration state where it initializes all of the hardware and reads the configuration file from the SD card. It then waits for the encoder to be homed by moving the index mark on the glass scale past the read head. After startup, it enters an idle state from which an experiment can be initiated or the tuning state entered. The sequence after starting an experiment is discussed in the experimental procedure section. The tuning state offers a serial command shell which exposes all of the CyberDiver functionality for testing and tuning. The shell maintains circular character buffers for input commands or output data, and low level plain text parsing routines. Individual subsystems of the control software can implement their own handlers for serial commands and register them with the shell, thereby creating an extensible and modular system. The control board enumerates as a virtual communication port when plugged in via USB, so serial commands can be passed to the CyberDiver using a serial terminal or a host application in Python or MATLAB, for instance. In order to tune the control loop gains of the system, we utilize a MATLAB script that loads new gains into the controller via the serial interface, commands a step change in the set point, and requests the logged data to be plotted and analyzed. As a result the control loop gains can quickly be iteratively tuned until achieving the desired step response. A list of all the CyberDiver serial commands is available in the following section which serves as documentation for the complete set of functionalities of the device.

While the active control is running and the system is taking measurements, the control board samples the accelerometer readings based on interrupts triggered by a data ready signal from the IMU chip. When the IMU chip requests data to be read, the same fast SPI routine is used to retrieve the data. The quadrature input signal from the encoder is managed by hardware so the nose displacement is always available in the appropriate timer count register. With sensor readings taken care of, the procedure during the time critical update function on the control board is to calculate the desired coil current depending on the operating mode, communicate this to the power board, and log data to the external RAM which eventually finds its way to the SD card in the main loop update. The controller can operate in current control mode, force control mode, position control mode, or simulated structure mode. In current control mode the commanded value is simply passed directly through to the power board. In force control mode the commanded force is converted to a current based on the measured displacement and the coil force coefficient calibration curve. In position control mode the control board runs its own PID loop on the nose displacement to decide what current to command from the power board. Finally, in simulated structure mode the target force is computed based on the configured structural curve as shown in Figure \ref{fig:fig2}(D). The performance of the controller calculations is profiled by toggling a spare MCU general purpose input-output (GPIO) pin at the start and end of the calculation and measuring the interval with a logic analyzer. The selected update rate for the controller is 50 kHz. This value is constrained by the coil current measurements rather than the CPU speed; a current measurement is corrupted if it occurs too close to an edge of the PWM due to the high rate of change of the common mode voltage at the sense amplifier. Beyond 50 kHz PWM frequency, noise in the current measurement from the common mode transients increases substantially; the PWM frequency sets the upper bound on the control loop update rate since the PWM duty cycle can only be changed once per PWM period. Additionally, power dissipation and heating in the H-bridge MOSFETs increases with PWM frequency since more time is spent switching. An alternative solution would be to use a linear power amplifier such as OPA549 as in other work \cite{lee2022mnn}. The linear amplifier regulates the coil current by varying an internal resistor so would produce no switching common mode transients. However, this system would require the generation of a high-current negative power rail. The techniques we use to improve the current sensing in the presence of switching transients are discussed in a later section in more detail. In addition to those strategies, we briefly explored the use of finite impulse response digital filtering or Kalman filter state estimators to improve the noise level in the current measurement. No discernible benefits to the overall system performance were found with these additions, likely because reducing the noise with computational techniques introduces some delay in the measurement which is equally detrimental to the control loop. In spite of the limitations from the coil switching transients in the current measurements, the overall system performance is adequate to implement a cyber-physical system on the fast time scales of water entry as demonstrated in our validation data sets.

\subsubsection*{CyberDiver serial commands} \label{sec:cyberdiver_serial_commands}

The list of serial commands for the CyberDiver is separated into categories by subsystem. Arguments for the commands and their types are indicated with parentheses: (argument:type). The CyberDiver will respond first with “ack” after successfully processing a command, and then with the requested value(s) if calling a command that returns data. All communications with the CyberDiver are ASCII-encoded and terminated with the line feed (LF) character. 

\paragraph{Connection:}

\begin{description}
    \item[connect] If the CyberDiver is currently in the idle state (\textit{i.e.} not running an experiment), this will cause it to enter the tuning state where it responds to serial commands.
    \item[disconnect] If the CyberDiver is in the tuning state, send this command to return to the idle state after which an experiment can be started by pressing the user button. This command must be sent before disconnecting the USB cable or else the CyberDiver will be stuck in the tuning state. Alternatively, power cycle the device after saving the configuration parameters to return to the idle state.
\end{description}

\paragraph{Controller:}

\begin{description}
    \item[controller set mode (mode:str)] Sets the operating mode of the CyberDiver controller from one of five choices. In ``idle'' mode the controller does nothing, and commands zero current from the power board. In “current\_control” mode the controller tries to maintain a set point coil current. In “force\_control” mode the controller tries to maintain the total coupling force at a set point value. In “position\_control” mode the controller tries to maintain a prescribed nose displacement. In “simulated\_structure” mode the controller operates as a cyber-physical system and tries to simulate the programmed structural curve.
    \item[controller get mode] Returns the string for the current operating mode of the CyberDiver controller.
    \item[controller set setpoint (setpoint:float)] Specifies the set point value of the controller. The units depend on the operating mode (amps, Newtons, or mm). The set point is ignored in the idle and simulated structure modes.
    \item[controller get setpoint] Returns the floating point set point value of the controller whose units depend on the controller operating mode.
    \item[controller set led (led:int)] Turns on or off the blue indicator LED on the controller board depending on whether the argument is 0 or 1.
    \item[controller get led] Returns the state of the indicator LED on the power board.
    \item[controller set state (mode:str) (setpoint:float) (led:int)] Combined command to set the operating mode, controller set point, and LED state simultaneously.
    \item[controller get state] Combined command that returns all of the state parameters of the controller (operating mode, set point, and LED state).
    \item[controller get position] Returns the measured encoder position in mm.
    \item[controller get velocity] Returns the measured encoder displacement velocity in mm/s.
    \item[controller get current] Returns the measured coil current in amps.
    \item[controller get force] Returns the calculated total coupling force between the nose and body in Newtons.
    \item[controller get duty] Returns the voice coil duty cycle as a floating point value between -1 and 1.
    \item[controller get accel] Returns the measured axial acceleration of the impactor body in g.
    \item[controller get output] Returns the output of the control board calculation. This is the current in amps which is sent to the power board as the set point for its coil current PID loop.
    \item[controller get data] Gets the latest data point from the controller, which contains all of the parameters that are logged during an experiment. These include (in order) the time stamp in microseconds, the measured current, the calculated total coupling force, the position of the encoder, the estimated velocity of the encoder, the coil duty cycle, the x, y, and z components of measured body acceleration, the controller set point, the controller mode string, and the indicator LED state.
    \item[controller resettime] Resets the hardware timer count for the controller time stamps to zero.
    \item[controller sequence start] Starts the currently programmed experimental sequence. The sequence specifies the operating modes, set points and times to transition between them during an experiment.  
    \item[controller sequence stop] Stops running the experimental sequence.
    \item[controller sequence clear] Clears the experimental sequence. This command will set the sequence length to zero and disable looping. Each step in the memory buffer will be set to idle state with a set point of zero. 
    \item[controller sequence set length (length:int)] Sets the number of steps in the experimental sequence.
    \item[controller sequence get length] Returns the number of steps in the experimental sequence.
    \item[controller sequence set looping (looping:int)] Sets whether the experimental sequence should stop once it reaches the end or loop depending on an argument of 0 or 1. If the sequence is not set to loop, the controller mode will return to idle once the sequence is complete.
    \item[controller sequence get looping] Returns whether or not the experimental sequence is set to loop.
    \item[controller sequence set step (index:int) (time:int) (mode:str) (setpoint:float) (led:int)]~\\ Program a step in the experimental sequence. The index argument specifies which step will be set and the time argument specifies how much time to spend in that step in microseconds. The other arguments specify what action to perform in that step of the sequence by dictating the controller state as described previously.
    \item[controller sequence get step (index:int)] Returns the information associated with a given step in the experimental sequence specified by the index argument. The information returned is in a format identical to the last four arguments of the previous command.
    \item[controller config set gains (P:float) (I:float) (D:float) (FF:float)] Configures the gains of the position PID controller, including the feed-forward (\textit{i.e.} open loop) gain FF. The output of the position PID controller dictates the total restoring force that the coil plus passive structure should apply to correct the position error. The position error in the PID controller implementation has units of mm.
    \item[controller config get gains] Returns the gains of the position PID controller in the same format and order as the previous command.
    \item[controller config set tau (tau:float)] Sets the time constant of the low pass filter on the derivative term of the PID controller in units of seconds.
    \item[controller config get tau] Returns the time constant of the derivative term low pass filter in seconds.
    \item[controller config set period (T:float)] Sets the update period $T$ of the control board calculation in seconds (\textit{i.e.} how frequently a new target coil current is calculated and passed to the power board). This value should only be adjusted with a logic analyzer connected to verify that the microcontroller can handle the requested update period. This value is also the period of the PID controller when running in position control mode as well as the base data recording rate.
    \item[controller config get period] Returns the update period of the control board calculation in seconds.
    \item[controller config set limits (min:float) (max:float)] Sets the output limits of the position PID controller in units of Newtons.
    \item[controller config get limits] Returns the minimum and maximum output limits of the position PID controller in Newtons.
    \item[{\parbox[c]{\linewidth}{\vspace{2ex}controller config set calibration coil (order:int) (min:float) (max:float) (fastcompute:int) (coeff0:float) (coeff1:float) \ldots}}]~\vspace{2ex}~\\Sets the calibration polynomial for the coil force coefficient. The input will be the displacement $\delta$ in mm and the output is the force coefficient $k_f(\delta)$ in Newtons per amp. The coefficients should be provided in ascending order, where the contribution of the nth coefficient is $a_n \delta^n$. The order argument specifies the polynomial order and should agree with the number of coefficients provided. The fast compute argument (0 or 1) tells the controller whether to directly evaluate the polynomial at each update or whether to pre-compute a lookup table. The minimum and maximum arguments impose limits on the input to the evaluated polynomial. Input values are clamped to the specified limits before performing the evaluation.
    \item[controller config get calibration coil] Returns the calibration polynomial for the coil force coefficient in the same format as the previous command.
    \item[{\parbox[c]{\linewidth}{\vspace{2ex}controller config set calibration passive (order:int) (min:float) (max:float)\\(fastcompute:int) (coeff0:float) (coeff1:float) \ldots}}]~\vspace{2ex}~\\Sets the calibration polynomial for the passive structural response of the flexure bearings. The input to the polynomial is the displacement in mm and the output is the restoring force in Newtons. The description of the arguments is identical to the previous calibration commands.
    \item[controller config get calibration passive] Returns the calibration polynomial for the passive structural response of the flexure bearings in the same format as the previous command.
    \item[{\parbox[c]{\linewidth}{\vspace{2ex}controller config set calibration structure (order:int) (min:float) (max:float)\\(fastcompute:int) (coeff0:float) (coeff1:float) \ldots}}]~\vspace{2ex}~\\Sets the calibration polynomial for the simulated structural curve representing the target behavior of the cyber-physical system. This curve is used when the controller is in the simulated structure operating mode (in this mode, the set point is ignored). The input to the polynomial is the displacement in mm and the output is the desired cyber-physical restoring force in Newtons. The description of the arguments is identical to the previous calibration commands.
    \item[controller config get calibration structure] Returns the calibration polynomial for the simulated structural curve representing the target behavior of the cyber-physical system in the same format as the previous command.
    \item[controller config set damping (damping:float)] Sets the linear damping coefficient to use in the simulated structure mode in N/(mm/s).
    \item[controller config get damping] Returns the linear damping coefficient used in the simulated structure mode in N/(mm/s).
\end{description}

\paragraph{Power board:}

\begin{description}
    \item[powerboard set mode (mode:str)] Sets the operating mode of the power board to one of two options. The code implements a 1 ms delay before and after switching the power board mode to ensure that the SPI communication is not corrupted. The ``idle'' mode disables the H-bridge and waits for the power board to receive configuration parameters over SPI. The ``running'' mode enables the H-bridge and updates the coil control PID loop. In this state the power board expects to receive current set point values over SPI and will respond with the latest measurement of the coil current. When the power board is in the running mode, the physical button can be used to disable the power stage (\textit{i.e.} an emergency stop if the control loops become unstable while tuning). However, the stop button state is not transmitted back to the control board. The red LED on the power board is illuminated when the power stage is enabled.
    \item[powerboard get mode] Returns the operating mode of the power board.
    \item[powerboard set current (current:float)] Sends a current set point value (in amps) to the power board. This command bypasses the outer control loop so the relative velocity will not be sent to the power board for back-EMF rejection as it would be in normal operation.
    \item[powerboard get current] Returns the latest coil current measured by the power board in amps.
    \item[powerboard get duty] Returns the latest duty cycle output by the power stage as a floating point value between -1 and 1.
    \item[powerboard config set gains (P:float) (I:float) (D:float) (FF:float)] Configures the gains of the coil current PID controller, including the feed-forward (\textit{i.e.} open loop) gain FF. The gains can only be set when the power board is in the idle state; otherwise this command will be ignored. The input to the PID controller is the coil current error in amps and the output is the duty cycle as a floating point value between -1 and 1 (proportional to the voltage applied to the coil, nominally between -25.2 V and 25.2 V).
    \item[powerboard config get gains] Returns the gains used in the coil current control loop in the same order and format as the previous command.
    \item[powerboard config set period (T:float)] Sets the update period $T$ of the coil controller PID loop in seconds. This value should only be adjusted with a logic analyzer connected to verify that the microcontroller can handle the requested update period. The period can only be set when the power board is in the idle state; otherwise this command will be ignored.
    \item[powerboard config get period] Returns the update period of the coil control PID loop in seconds.
    \item[powerboard config set tau (tau:float)] Sets the low-pass time constant of the derivative term in seconds. The time constant can only be set when the power board is in the idle state; otherwise this command will be ignored.
    \item[powerboard config get tau] Returns the low-pass time constant of the derivative term in seconds.
    \item[powerboard config set kffvel (FF:float)] Sets the velocity feed-forward gain of the coil current control loop. This is an additional term in the control used to proactively reject back-EMF due to relative motion between the coil and the magnetic field assembly by applying an additional duty cycle contribution proportional to the velocity. This gain can only be set when the power board is in the idle state; otherwise this command will be ignored. We set this term to zero in our experiments since we did not find it necessary.
    \item[powerboard config get kffvel] Returns the velocity feed-forward gain of the coil current control loop.
\end{description}

\paragraph{Logger:}

\begin{description}
    \item[logger start] Tells the logging module of the control board code to start storing data in the circular buffer.
    \item[logger stop] Stops the logger from adding any more data to its buffer. This command freezes the contents of the data buffer so they can be accessed later.
    \item[logger stream start] Tells the logger to start streaming data over serial. This means any available data packets will immediately be encoded and sent over USB. Each data transmission begins by stating the number of data points in the packet. Then the data packets follow, with each packet using the same format as the ``controller get data'' command.
    \item[logger stream stop] Stops the data stream from the logger.
    \item[logger get packet] When the logger is not streaming, this command can be used to extract data packets at will. It returns the oldest data packet  in the logger circular buffer (following a first-in-first-out scheme). This command might return less than a complete packet if that is all that remains in the buffer.
    \item[logger get available] Returns the number of complete packets available in the data buffer.
    \item[logger set decimation (decimation:int)] Sets the decimation of the logger. A setting of 0 records a data point at each update of the controller; a setting of 1 skips every other data point and so on.
    \item[logger get decimation] Returns the decimation setting of the logger.
    \item[logger set location (location:str)] The logger module can buffer data in either the local MCU RAM (faster access) or in the 16 MB external SDRAM chip (more space). Arguments of ``mcu'' or ``sdram'' can be used to select between the two options. We almost exclusively use the SDRAM while tuning the CyberDiver or performing experiments.
    \item[logger get location] Returns the string corresponding to the memory location of the logger data buffer.
    \item[logger set packetsize (size:int)] Sets the number of data points that makes up a packet. 
    \item[logger get packetsize] Returns the size of a packet in data points.
\end{description}

\paragraph{Encoder:}

\begin{description}
    \item[encoder home] Waits for the encoder count to be reset by moving the encoder index past the optical window.
    \item[encoder set offset (offset:float)] Sets the offset between the index mark and the neutral position of the flexure bearings in mm.
    \item[encoder get offset] Returns the offset between the index mark and the neutral position of the flexure bearings in mm.
    \item[encoder get counts] Returns the position of the encoder in counts which each represent 1 micron of displacement.
    \item[encoder get position] Returns the position of the encoder in mm with the offset taken into account. 
\end{description}

\paragraph{Accelerometer:}

\begin{description}
    \item[accel set range (range:int)] Sets the range of the accelerometer (IIS3DWB) in g. The allowed values are 2, 4, 8 and 16. Other values will be ignored. A setting of 2 configures the accelerometer to measure in the range between -2 g and 2 g.
    \item[accel get range] Returns the configured range of the accelerometer with the same interpretation as the previous command.
    \item[accel set lpf (lpf:int)] Configures the low-pass filter internal to the accelerometer chip. The allowed values are 4, 10, 20, 45, 100, 200, 400, and 800. A setting of 4, for example, means the LPF is configured such that the output bandwidth is one quarter of the sampling rate (26.6 kHz). 
    \item[accel get lpf] Returns the configuration of the low-pass filter with the same interpretation as the previous command.
    \item[accel set offsets (x:float) (y:float) (z:float)] Sets the DC offsets of each axis of the accelerometer in g. These values are applied internally in the sensor hardware. The offsets are caused due to the mechanical stresses during the soldering process and thus may vary between each board. 
    \item[accel get offsets] Returns the configured DC offsets of each axis of the accelerometer in g.
    \item[accel get data] Returns the most recent acceleration values measured by the sensor for the x, y, and z axes in g. The axis directions are indicated on the control board silkscreen layer.
\end{description}

\paragraph{Configuration:}

\begin{description}
    \item[config save] Saves all of the current configuration parameters to a file on the SD card. The global configuration struct is saved directly as binary. When the CyberDiver is first turned on, it loads all of the configuration parameters from the SD card.
    \item[config load] Loads the configuration parameters from the file on the SD card. This can be useful to restore the configuration if it reaches an unintended state while tuning.
    \item[config get version] Returns the firmware version.
    \item[config set state idle (mode:str) (setpoint:float) (led:int)] Configures the controller state when the CyberDiver first initializes and is not running an experiment, using the same format as the ``controller set state'' command.
    \item[config get state idle] Returns the controller state when the CyberDiver first initializes and is not running an experiment, using the same format as the previous command.
    \item[config set state armed (mode:str) (setpoint:float) (led:int)] Sets the controller state while the CyberDiver is hanging from the magnet, ready to be dropped, using the same format as the ``controller set state'' command.
    \item[config get state armed] Returns the controller state when the CyberDiver is hanging from the magnet, using the same format as the previous command.
    \item[config set time staging (time:int)] Sets the time delay in microseconds after pressing the button to start an experiment before the armed mode is entered. This period allows the CyberDiver to be handled and attached to the magnet without inadvertently experiencing an acceleration trigger.
    \item[config get time staging] Returns the staging time delay in microseconds.
    \item[config set time running (time:int)] Sets the total time to run an experiment and record data in microseconds. The CyberDiver returns to the idle state after the run time elapses.
    \item[config get time running] Returns the experiment run time in microseconds.
    \item[config set trigger (accel:float)] Sets the acceleration threshold value in g to trigger the transition from armed to running. When the absolute value of the measured acceleration goes below this threshold the experiment starts. This is used to detect free-fall in our setup.
    \item[config get trigger] Returns the acceleration trigger threshold value in g.
\end{description}

\subsubsection*{H-bridge modulation scheme} \label{sec:hbridge}

\begin{figure}
    \centering
    \includegraphics[width = \linewidth]{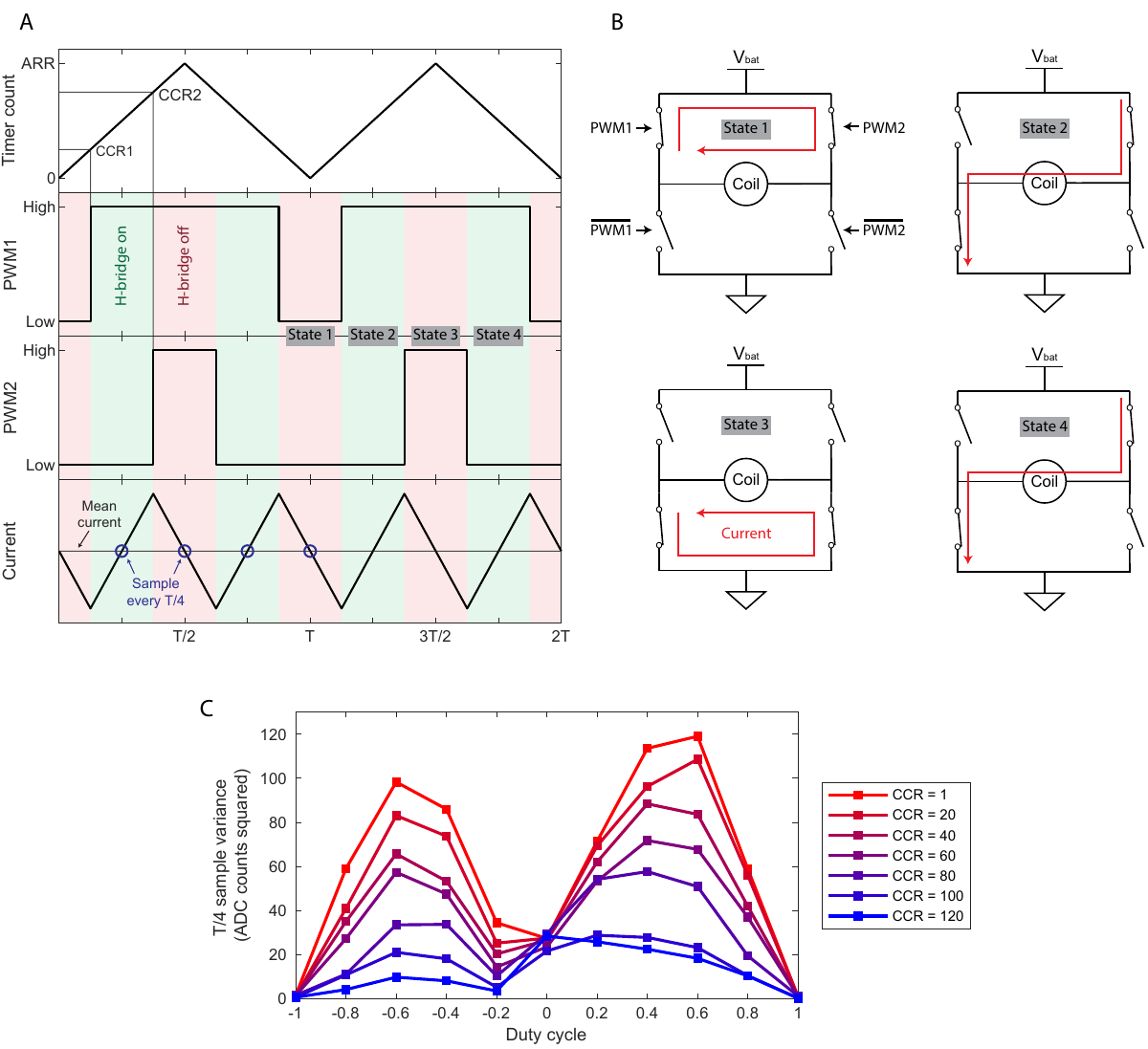}
    \caption{\setlength{\baselineskip}{12pt} \textbf{The H-bridge is modulated using a four-state scheme in order to improve efficiency and current measurement noise.} (\textbf{A}) The configuration of the two center-aligned PWM channels and their response to the timer count over two periods are illustrated. In addition, the current in the coil is shown assuming that the PWM frequency is fast enough compared to the coil's electrical time constant that the coil's response appears linear rather than exponential. By sampling the coil current every $T/4$ interval, each of the samples is equal to the mean value of the current assuming that the system is in steady state. Additionally, samples at this interval are guaranteed to be as far from the PWM edges as possible even as the timer compare registers (CCR1 and CCR2) are varied. (\textbf{B}) Diagrams of the four states of the H-bridge produced by the modulation scheme are shown with illustrations of the coil current. (\textbf{C}) We tune the current sampling phase to account for delays introduced by the sense amplifier, ADC conversion, and low-pass filter in order to minimize the variance between the four samples in each period. Each CCR count in the legend corresponds to a time delay of 11.9 ns.}
    \label{fig:fig9}
\end{figure}

The scheme implemented to modulate the H-bridge is illustrated in Figure \ref{fig:fig9}(A). The source for the PWM signals is a hardware timer operating in center-aligned mode -- at each clock pulse the timer counts up from zero and then back down once it reaches the configured value of the reload register (ARR). Hence the timer count forms a triangular wave form with period $T$ which in our implementation is 20 $\mu$s. The timer count is used to generate PWM signals in two channels. Each side of the H-bridge is actuated by one of the PWM channels via the gate driver circuitry which also inverts the actuation of the low side compared to the high side MOSFET as shown in Figure \ref{fig:fig9}(B). The duty cycles of the PWM channels are dictated by the values in the compare registers CCR1 and CCR2 corresponding to each channel. In the example shown, the PWM signal is high when the timer count exceeds the compare register value and low when the timer count is less than the compare register value. These values can be changed by software on the fly at the end of each period in order to adjust the actuation effort of the H-bridge. In each period, the H-bridge traverses four states as shown in Figure \ref{fig:fig9}(B). States 2 and 4 are active states in which the battery applies a voltage across the coil in order to drive the coil current. States 1 and 3 are passive states in which the coil current recirculates in either the upper or lower side of the bridge but the battery applies no voltage to the coil. One benefit of four state modulation as shown, and the motivation for distinct states 1 and 3, is that it evenly distributes heating associated with switching losses between all MOSFETs in the circuit. The fraction of time spent in each state dictates the overall actuation level of the H-bridge. For instance, if CCR1 = CCR2 then the H-bridge will spend all of the period in passive states 1 and 3 and no voltage will be applied from the battery. A case where CCR1 = 0 and CCR2 = ARR represents maximum actuation in which the battery voltage is applied to the coil for the entire period. By then swapping the values in CCR1 and CCR2 we can achieve the case of maximum actuation in the other direction. If we let $D$ represent the overall actuation level, or duty cycle, of the H-bridge ranging from -1 to 1, then the values of the compare registers are computed as
\begin{equation}
    \textrm{CCR1} = \frac{1}{2} \textrm{ARR} (1 + D), \hspace{1cm} \textrm{CCR2} = \frac{1}{2} \textrm{ARR} (1 - D).
\end{equation}
By maintaining symmetry in the values of the compare registers, current samples captured every quarter period $T/4$ will be centered between PWM edges for intermediate bridge duty cycle values as shown in Figure \ref{fig:fig9}(A). The current sense ADC conversions are triggered by a separate hardware timer so it is convenient to be able to sample at regular intervals while also avoiding common mode transients associated with PWM edges in the current measurement as much as possible. Due to the design of the modulation scheme, the switching frequency in the coil current is double the timer frequency, reducing the overall current ripple. Additionally, assuming that the switching period is much smaller than the coil inductive time constant (so that the current wave form is approximately triangular) and that the system has reached steady state in terms of the mean current, the current samples at every $T/4$ interval will be equal to the mean current. The four samples per period are averaged to further reduce noise in the measurement and the average value is used in the control loop calculation to update the duty cycle for the next period. At some duty cycles, the PWM edges necessarily land close to the current sample times; we omit a sample from the period average if it lands within 1.25 $\mu$s of a PWM edge to mitigate the error influence of common mode transients. Finally, we account for the lag of the current sampling circuitry by slightly shifting the phase of the current samples until the variance between the four samples per period is minimized as shown in Figure \ref{fig:fig9}(C). Sources of lag in the current sense circuitry include the differential amplifier, the low-pass filter, and the conversion time of the successive approximation ADC. The overall lag will manifest as a shifting of each of the current samples along the triangle waveform and away from the mean. Here, the CCR value corresponds to the compare register of the timer which is used to trigger ADC conversions. We use the value CCR = 100 in order to minimize the sampling variance which corresponds to delaying the current sample times by 1.19 $\mu$s relative to the PWM timer. The asymmetry in the sample variance result likely stems from the fact that the shunt resistor is placed to one side of the coil in the H-bridge, meaning that the common mode voltage at the current sense amplifier is different depending on whether the duty cycle is positive or negative.      

\subsubsection*{PID controller}

The CyberDiver utilizes two PID controllers to regulate the coil current and the nose displacement. The coil current PID controller is the inner loop which runs on the power board microcontroller and receives set point commands from the control board. The coil current PID controller is active in all operating modes of the CyberDiver; the nose displacement PID controller is only used in the position control mode. Both of the PID controllers are implemented in the same way. The continuous time domain transfer function of the PID controller $C(s)$ relates the control signal $\bar{u}(s)$ and the measured error $\bar{e}(s)$ according to
\begin{equation}
    C(s) = \frac{\bar{u}(s)}{\bar{e}(s)} = K_p + K_i \frac{1}{s} + K_d \frac{s}{s \tau + 1}
\end{equation}
where $K_p$, $K_i$ and $K_d$ are the gains of the proportional, integral and derivative terms respectively. A low-pass filter with time constant $\tau$ is included in the derivative term to mitigate the effects of high-frequency noise in the measured error which could contribute to controller instability if left unchecked. The continuous-time transfer function is converted to discrete-time difference equations that can be implemented on a digital microcontroller through the bilinear transformation
\begin{equation}
    s \rightarrow \frac{2}{T} \frac{z-1}{z+1}
\end{equation}
where $T$ is the discrete controller period \cite{ogata2009modern}. Subsequently the inverse Z-transform yields the difference equations used to implement each of the controller terms including the proportional term
\begin{equation}
    p[n] = K_p e[n],
\end{equation}
integral term
\begin{equation}
    i[n] = \frac{K_i T}{2} \left(e[n] + e[n-1]\right) + i[n-1],
\end{equation}
and derivative term
\begin{equation}
    d[n] = \frac{2 K_d}{2 \tau + T} \left( e[n] - e[n-1] \right) + \frac{2 \tau -T}{2 \tau + T} d[n-1].
\end{equation}
For greater flexibility in the software controller, we also introduce a feed-forward term 
\begin{equation}
    ff[n] = K_{ff} s[n]
\end{equation}
which depends only on the set point $s[n]$. By setting $K_{ff} = 0$, one can recover the pure PID controller. In these terms, $e[n]$ refers to the current sample of the measured error. The full controller is given by
\begin{equation}
    u[n] = ff[n] + p[n] + i[n] + d[n].
\end{equation}

Some aspects of the controller are further modified to improve the performance of the practical implementation. Configurable maximum and minimum limits are imposed on the output of the controller corresponding to the physical limits of the system. Anti-windup is performed by clamping the value of the integral term to the difference between the controller output limit and the current value of the proportional term. Finally, the derivative term is modified to use only the measured value rather than the error value since otherwise the derivative term would produce an undesirable ``kick'' in response to a step change in the set point -- in our dynamic system the set point for the PID controllers tends to change rapidly.  

\subsubsection*{Force test machine}

\begin{figure}
    \centering
    \includegraphics[width = 0.9\textwidth]{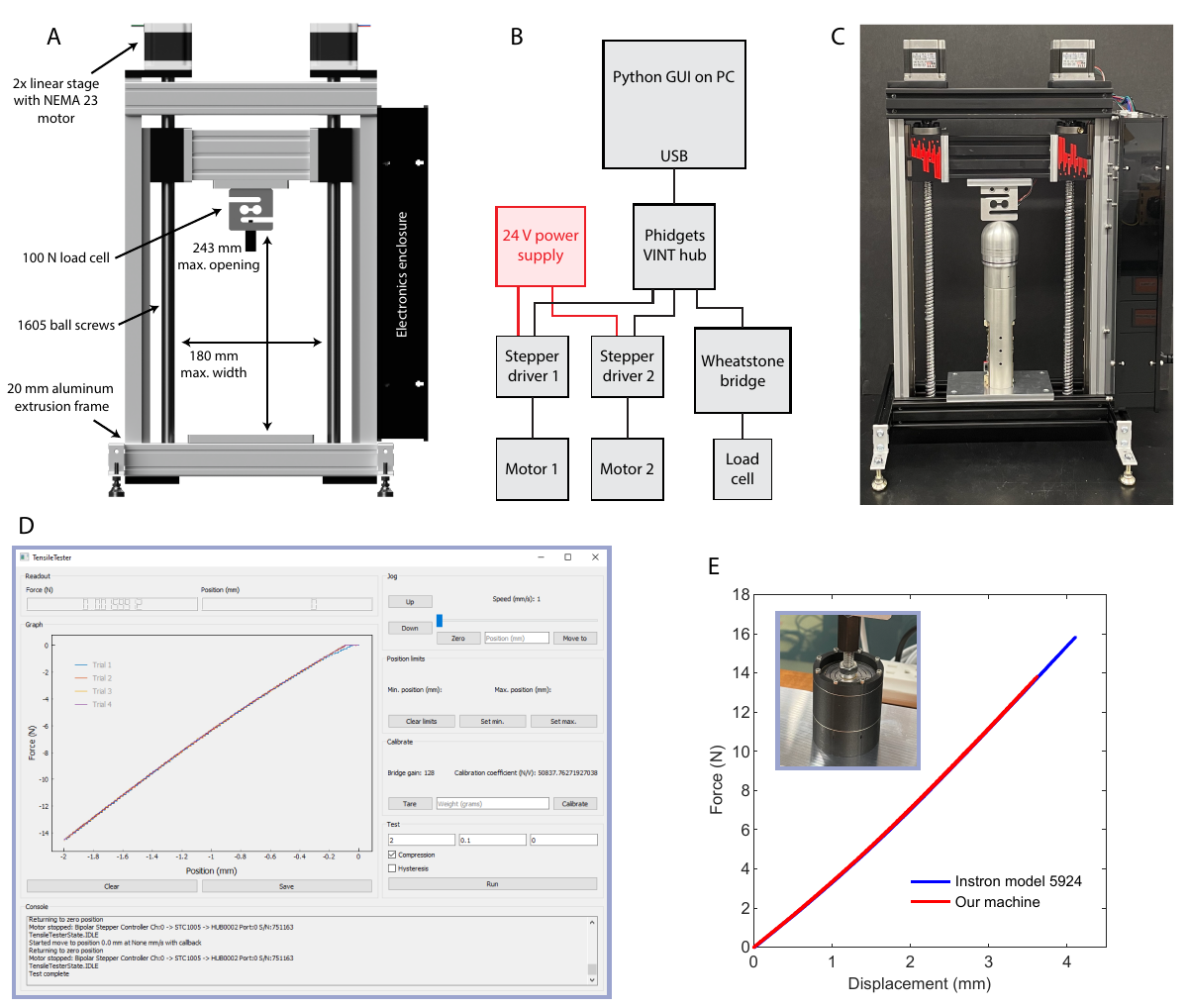}
    \caption{\setlength{\baselineskip}{12pt} \textbf{Custom force testing fixture for calibrating and tuning the CyberDiver.} (\textbf{A}) A computer-aided-design render of the force testing machine is shown with the relevant dimensions and components labeled. (\textbf{B}) A Python GUI on a host computer is used to interface with a Phidgets VINT hub, control the force testing sequence, and acquire data. The hub interfaces with a Wheatstone bridge module to read the load cell and stepper drivers which position each of the linear stages. (\textbf{C}) The CyberDiver can be mounted in the testing machine to measure its force-versus-displacement response. (\textbf{D}) For simple force tests, a graphical user interface was created using Python, Qt, and the Phidgets API. The extents and direction of the force test can be configured and each trial's data is plotted in real time as it is gathered. (\textbf{E}) Results of a compression test using our machine are compared to a test of the same specimen in an Instron model 5924 universal force testing machine. The test specimen is a prototype disk flexure bearing as shown in the inset photograph.}
    \label{fig:fig10}
\end{figure}

A custom force testing machine was developed for use in calibrating and characterizing the CyberDiver as shown in Figure \ref{fig:fig10}, motivated by the need to perform automated testing in which both the test machine and the CyberDiver are controlled simultaneously by a script on a host computer. The force testing machine uses a bridge-style construction in which the test head is affixed to the columns on both ends for increased rigidity as shown in Figure \ref{fig:fig10}(A). The machine frame is assembled from 20 mm series aluminum extrusion and a limited number of custom parts machined from 10 mm thick MIC6 aluminum plate. MIC6 aluminum plates are stress-relieved, easy to machine, and available with tight thickness tolerances, resulting in an affordable and quick method of making custom metal parts. The parts for the force test machine can be machined using only a drill press and band saw, common low-precision tools, yet retain the tight parallelism tolerance of the MIC6 stock. The columns of the machine are comprised of two commercially available linear axes (Fuyu FSL40) with ball bearing linear guides, NEMA 23 stepper motors, and 16 mm diameter ball screws. The overall height of the machine is 565 mm and it supports sample sizes up to 243 mm in height and 180 mm in width, accommodating the fully assembled CyberDiver as shown in Figure \ref{fig:fig10}(C). A laser-cut acrylic enclosure is affixed to the side of the machine frame which houses the power supply, motor/load cell connectors, and control electronics. The Phidgets ecosystem is used to control the machine via their Versatile Interface (VINT) hub (Phidgets HUB0002\_0) with a USB connection to a host computer that runs the Python graphical user interface (GUI) or automated testing script. A schematic of the electronics is shown in Figure \ref{fig:fig10}(B). Force on the specimen is measured using a 100 N load cell (Phidgets FRC4161\_0) connected to a Wheatstone bridge with 24 bit ADC (Phidgets DAQ1500\_0). Two stepper drivers with 4 A capacity (Phidgets STC1005\_0) are used to synchronously position the two linear axes of the machine and apply a controlled displacement to the sample. Given the ball screw pitch and stepper driver micro-stepping capabilities, the theoretical positioning resolution is 1.6 $\mu$m but the low-cost ball screws are unlikely to achieve this level of accuracy. The stepper drivers are powered with a 24 V DC power supply (Mean Well LRS-350-24). A graphical user interface as shown in Figure \ref{fig:fig10}(D) was developed to control the machine using the PyQt5 library for UI elements and the Phidgets Python API for interfacing with the hardware. The interface allows the user to jog the machine with adjustable speed, set the home position, calibrate the load cell using calibration weights, tare the load cell, and perform simple test procedures. Simple compression or tension tests can be run in which the user selects the displacement extent, displacement rate, maximum allowable force, and whether to also record the load cell on the return trip to characterize hysteresis in the sample. As each test is performed, the measurements are plotted in real time on the force-versus-displacement graph. Once the testing has been completed, the results may be exported to a CSV file for further processing. The GUI is used to characterize the CyberDiver while it runs in a linear stiffness cyber-physical operating mode. When calibrating the device, the Phidgets API may be used directly in a Python script to control the force testing machine programmatically. The performance of the custom force testing machine was validated by comparing its measurements to those taken by a commerical force testing machine (Instron model 5924 with 500 N load cell) with the results shown in Figure \ref{fig:fig10}(E). Using a prototype disk flexure bearing as the specimen as shown in the inset photograph, the measurement from our custom force testing machine agrees excellently with the commercial machine. Additionally, we measured the stiffness of our machine frame to be 402 N/mm by performing a compression test in which the sample is omitted and the load cell is compressed directly against the machine base plate. 

\subsubsection*{Error analysis}

We prescribe the impact speed for each water entry experiment by calculating the drop height $H$ according to
\begin{equation}
    V = \sqrt{2 g H},
\end{equation}
marking a string with the appropriate distance, and suspending it from the electromagnet. Then the height of the drop arm is adjusted until the mark coincides with the water surface. Each time we adjust the drop arm height, we take a high-speed video of impact to ensure that the target impact speed is achieved. When calculating the contribution of uncertainty in the impact speed to the overall experimental error, we assume that the standard deviation of the drop height $\sigma_H$ is 10 mm. Then the propagated error in the impact speed can be calculated as
\begin{equation}
   \sigma_V = g (2 g H)^{-1/2} \sigma_H = \frac{g}{V} \sigma_H.
\end{equation}

In analysis involving the impactor stiffness, such as the computation of the hydroelastic number, we use the nominal programmed stiffness values since the cyber-physical system tends to perform accurately. The uncertainty in the cyber-physical stiffness value is taken as the maximum deviation between the nominal value and either the quasi-static or dynamic method of measuring the stiffness, as reported in Table \ref{tab:stiffness_error}. Since the CyberDiver experimental data recording is triggered by detecting free fall, the experimental data is very repeatable in time and requires no further alignment between trials. The estimate for the peak acceleration in each experimental case is obtained as the maximum value of the ensemble average of the acceleration traces from the five experimental replicates. The reported standard deviation of the maximum acceleration $\sigma_{a_m}$ is the standard deviation between the acceleration traces at the time of the peak. The uncertainty in the maximum acceleration tends to be small because the experiments are very repeatable. However, unintended higher frequency structural oscillations in the acceleration trace contribute some error that is not captured by this method. Other parameters used in analyzing the experimental data include the total mass and mass ratio of the impactor, its radius, and the water's density. These values can be measured directly and with high precision, and we neglect their contribution to the error.

The uncertainty $\sigma_{R_F}$ on the hydroelastic number $R_F$ is thus calculated as

\begin{equation}
    R_F = \sqrt{\frac{k}{\alpha(1-\alpha)M}}\frac{R}{V},
\end{equation}

\begin{equation}
    \frac{\partial R_F}{\partial V} = -\frac{1}{V} R_F,
\end{equation}

\begin{equation}
    \frac{\partial R_F}{\partial k} = \frac{1}{2} \frac{R}{V} \left( \frac{k}{\alpha (1-\alpha)M} \right)^{-1/2}\frac{1}{\alpha (1-\alpha)M} = \frac{1}{2k} R_F,
\end{equation}

\begin{equation}
    \sigma_{R_F} = \sqrt{\left(\frac{\partial R_F}{\partial V}\right)^2 \sigma_V^2 + \left(\frac{\partial R_F}{\partial k}\right)^2 \sigma_k^2} = R_F \sqrt{\left(\frac{\sigma_V}{V} \right)^2 + \left(\frac{\sigma_k}{2k} \right)^2}.
\end{equation}

The uncertainty $\sigma_{C_D}$ on the peak dimensionless force $C_D$ is calculated as

\begin{equation}
    C_D = \frac{M a_m}{\frac{1}{2} \rho V^2 \pi R^2},
\end{equation}

\begin{equation}
    \frac{\partial C_D}{\partial a_m} = \frac{1}{a_m} C_D,
\end{equation}

\begin{equation}
    \frac{\partial C_D}{\partial V} = -2 V^{-3} (C_D V^2) = \frac{-2}{V} C_D,
\end{equation}

\begin{equation}
    \sigma_{C_D} = \sqrt{\left(\frac{\partial C_D}{\partial V}\right)^2 \sigma_V^2 + \left(\frac{\partial C_D}{\partial a_\textrm{max}}\right)^2 \sigma_{a_m}^2} = C_D \sqrt{\left(\frac{2 \sigma_V}{V} \right)^2 + \left(\frac{\sigma_{a_m}}{a_m} \right)^2}.
\end{equation}


\subsection*{Supplementary Text}

\subsubsection*{Details of the added mass model for impact}\label{sec:cyberdiver_added_mass_model}

The dependence of the peak deceleration of the flexible impactor on the hydroelastic number $R_F$ can be reconciled through the application of a simple mathematical model for water entry that considers the force contribution from the rapidly rising fluid added mass during the early time slamming phase of water entry. The classic model was developed by Von Karman and applied to the rigid spherical geometry by Shiffman and Spencer \cite{von1929impact, shiffman1945sphere2}. We extended the model to the case of a flexible impactor in our prior work \cite{antolik2023shm}. The model derives the equation of motion for the two mass impactor system based on conservation of momentum of the system comprising the impactor nose and the virtual fluid added mass \cite{von1929impact, abrate2013slamming, shiffman1945sphere1, shiffman1945sphere2, shiffman1947lens}
\begin{equation}
    \frac{d}{dt}[\dot{x}_n (\alpha M + m(x_n)) ] = F_s (\delta, \dot{\delta})
\end{equation}
where $m(x_n)$ represents the virtual quantity of fluid added mass that must be accelerated by the impactor as it enters the water. We make the assumption that the added mass is a function of nose depth $x_n$ alone. The value of the added mass for a sphere is known from the literature \cite{shiffman1945sphere2}. An arbitrary function for the forcing due to the structural coupling can be captured in $F_s(\delta, \dot{\delta})$ which depends on the nose displacement and its rate. In the case of the linear damped harmonic oscillator system $F_s (\delta, \dot{\delta}) = k \delta + c \dot{\delta}$ where $c$ is the structural damping coefficient. Carrying out the derivative the equation of motion for the impactor nose is
\begin{equation} \label{eq:cyberdiver_added_mass_nose}
    \ddot{x}_n + \frac{1}{\alpha M + m(x_n)} \frac{dm(x_n)}{dx_n} \dot{x}_n^2 - \frac{F_s (\delta, \dot{\delta})}{\alpha M + m(x_n)} = 0
\end{equation}
and the equation of motion for the impactor body is simply due to the structural forcing
\begin{equation} \label{eq:cyberdiver_added_mass_body}
    \ddot{x}_b+\frac{F_s (\delta, \dot{\delta})}{(1-\alpha)M} = 0.
\end{equation}
These equations are integrated numerically using the RK4 technique with initial conditions $\delta(0) = 0$ and $\dot{x}_n(0) = \dot{x}_b(0) = V$ and compared to the undamped experimental results in Figure \ref{fig:fig3}(D). To generate the theory curve the added mass model is run for each $R_F$ value in the range and the peak acceleration from that simulated impact is stored. We use the nominal programmed stiffness value in the model and a damping ratio $\zeta$ of 0.03 since it is the central value of the range found in dynamic impulse physical testing for the nominally undamped system. The relationship between the damping ratio and the dimensional damping coefficient used in the model is $c = 2 \zeta \sqrt{kM\alpha (1-\alpha)}$. This same model is used to generate the theoretical results for the damped cases in Figure \ref{fig:fig4}(C).

\subsubsection*{Simulations of impacts with nonlinear structural coupling}

\begin{figure}
    \centering
    \includegraphics[width = \textwidth]{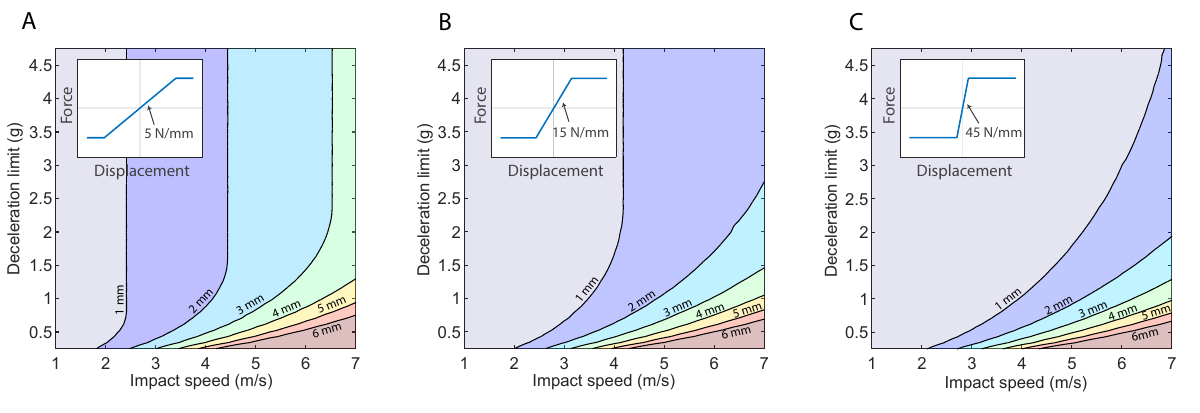}
    \caption{\setlength{\baselineskip}{12pt} \textbf{The two-way coupled added mass model is used to simulate the maximum displacement during impacts with the programmed nonlinear structural coupling.} Contours of maximum nose displacement are plotted as the impact speed and programmed deceleration limit (proportional to the force threshold) are varied. The stiffness of the linear region of the force-versus-displacement curve changes between 5 N/mm and 45 N/mm in (\textbf{A-C}), and less of the available nose travel is used when the linear region has higher stiffness. Thus a ``more strongly nonlinear'' structural coupling increases the operating envelope of the robotic diver given constrained nose travel.}
    \label{fig:fig11}
\end{figure}

With the goal of mitigating impact loading, an optimal structural coupling could be described as one that most effectively utilizes the available displacement while clamping the deceleration to the allowed value. Figure \ref{fig:fig11} shows contours of the maximum nose displacement for simulated impacts with varying speed and programmed force threshold as predicted by the added mass model. The stiffness of the linear region of the structural curve is varied between 5 N/mm to 45 N/mm in the three plots. As the stiffness of the linear region increases, the operating envelope of the diver expands because an impact at a given speed and acceleration limit tends to require less maximum travel. Given limits on both the peak deceleration and the available travel, the situation would further improve as the nonlinear conservative structural coupling approaches $F_s(\delta, \dot{\delta}) = (1-\alpha) M a_\textrm{lim} \textrm{sgn}(\delta)$ since this law always applies the maximum allowable restoring force to resist bottoming out. In practice the controller becomes unstable when trying to emulate infinite stiffness so this structural coupling is not possible to test.


\begin{table} 
    \centering
    \caption{\setlength{\baselineskip}{12pt} \textbf{Cyber-physical system linear stiffness error.} Target stiffness values for the CyberDiver when operating as a cyber-physical linear spring system and the corresponding stiffness values measured directly by compression testing and inferred from dynamic impulse response testing. The compression test result is the average of three compression trials and the impulse response result is based on the average frequency of five ring-down trials. For the sake of error propagation, the uncertainty is taken as the maximum deviation from the nominal value with either testing method.}
    \label{tab:stiffness_error}
    \begin{tabular}{lcccc}
        \\
        \hline
        Target & Compression test & Impulse test & $\sigma_k$ \\
        (N/mm) & (N/mm) & (N/mm) & (N/mm) \\
        \hline
        5 & 4.99 & 4.95 & 0.05 \\
        15 & 14.78 & 15.68 & 0.68 \\
        45 & 44.06 & 48.16 & 3.16 \\
        \hline
    \end{tabular}
\end{table}


\clearpage 

\subsection*{Captions for supplementary movies}

\paragraph{Caption for Movie S1.}
\textbf{CyberDiver entering the water at 4 m/s while behaving as a 5 N/mm undamped linear spring.}
The impact is filmed from above the water surface with front lighting to visualize the CyberDiver's construction, corresponding to Figure \ref{fig:fig1}(A). The playback is 100 times slower than actual speed. 

\paragraph{Caption for Movie S2.}
\textbf{Demonstration of CyberDiver actuation and its position control operating mode.}
As the CyberDiver hangs from a bungee, a demonstration of its position control operating mode is initiated by pressing the button on the control board. The nose then moves relative to the body in a programmed sinusoidal profile, illustrating the functionality of the actuator.

\paragraph{Caption for Movie S3.}
\textbf{Visualization of the impulse response of the cyber-physical system.}
An impulsive load is applied to the nose of the CyberDiver while it is suspended from a bungee and the vibrational response filmed. In the left frame, the cyber-physical system operates as a nominally undamped linear spring with 5 N/mm stiffness. In the right frame, the cyber-physical system operates as a nominally undamped linear spring with 15 N/mm stiffness. A corresponding visual difference between the amplitudes and frequencies of the vibrational responses is apparent. The playback is 25 times slower than actual speed.

\paragraph{Caption for Movie S4.}
\textbf{The CyberDiver manipulates the splash crown by performing active maneuvers during impact.}
The CyberDiver enters the water at 2 m/s and initiates an active maneuver at the moment of impact. Five different maneuver cases are shown where the impactor either extends or retracts the nose, resulting in amplification or attenuation of the splash formation, respectively. Movie version of Figure \ref{fig:fig6}(A-E).

\end{document}